\documentclass[12pt]{article}
\usepackage{xcolor}
\usepackage{hyperref}

\usepackage{amssymb}
\usepackage{amsmath}
\usepackage{amsfonts}
\usepackage{graphicx}
\usepackage{mathrsfs}
\usepackage{comment}
\usepackage{cite}
\usepackage{authblk}
\usepackage{array}
\usepackage{subcaption}

\newcommand{\Slash}[1]{{\ooalign{\hfil#1\hfil\crcr\raise.167ex\hbox{/}}}}
\newcommand{\bra}[1]{ \langle {#1} | }
\newcommand{\ket}[1]{ | {#1} \rangle }
\newcommand{\beq}{\begin{equation}}  \newcommand{\eeq}{\end{equation}}
\newcommand{\bef}{\begin{figure}}  \newcommand{\eef}{\end{figure}}
\newcommand{\bec}{\begin{center}}  \newcommand{\eec}{\end{center}}
\newcommand{\non}{\nonumber}  
\newcommand{\laq}[1]{\label{eq:#1}}  

\newcommand{\Eq}[1]{Eq.(\ref{eq:#1})}
\newcommand{\Eqs}[1]{Eqs.(\ref{eq:#1})}
\newcommand{\eq}[1]{(\ref{eq:#1})}
\newcommand{\Sec}[1]{Sec.\ref{chap:#1}}
\newcommand{\ab}[1]{\left|{#1}\right|}

\newcommand{\lac}[1]{\label{chap:#1}}

\def\({\left(}
\def\){\right)}

\def\diag{\mathop{\rm diag}\nolimits}

\def\O{\mathcal{O}}

\def\ebq{\end{equation} \begin{equation}}

\newcommand{\AND}{~{\rm and}~}
\newcommand{\EV}{ {\rm \, eV} }
\newcommand{\KEV}{ {\rm \, keV} }

\def\a{\alpha}
\def\b{\beta}

\def\d{\delta}
\def\e{\epsilon}
\def\f{\phi}
\def\g{\gamma}

\def\k{\kappa}

\def\m{\mu}

\def\D{\Delta}

\def\F{\Phi}

\def\tl{\tilde}
\def\*{\dagger}

\usepackage[normalem]{ulem}

\begin{document}
\begin{titlepage}
\begin{center}

\vspace{1.0cm}

{\Large\bf 
Quantum and Material Effects in Undulator-Based LSW Searches for Dark Photons
}

\vspace{1.0cm}

{\bf  Wen Yin}

\vspace{1.0cm}
{\em 
$^{1}${Department of Physics, Tokyo Metropolitan University, Tokyo 192-0397, Japan\\}
}

\vspace{1.0cm}
\abstract{ 
The dark photon is one of the simplest extensions of the Standard Model and provides a minimal laboratory for quantum-mechanical phenomena. Light-shining-through-a-wall (LSW) searches often adopt the dark photon-photon oscillation formula as if the sensitivity were independent of the light source, the wall, and the surrounding medium. In this paper, I revisit an LSW experiment whose light source is an undulator and systematically include various quantum effects: finite wave packets, kinematical suppression due to the microscopic structure of the source, and mixing suppression/enhancement in the wall and the air. We find that the resulting sensitivities deviate significantly from those obtained with the na\"{i}ve oscillation formula, especially depending on the mass of the dark photon, relevant to reflective index of the medium or walls, there can be resonance effects enhancing the sensitivity significantly. Accounting for these effects, we show that placing a photon detector outside the shielding along the beamline of a synchrotron facility enables an economical, parasitic LSW search for dark photons. 
}

\end{center}
\end{titlepage}

\setcounter{footnote}{0}
\section{Introduction}

One of the simplest extensions of the Standard Model of particle physics is the dark photon model~\cite{Holdom:1985ag}. 
The Lagrangian can be written as
\beq
\Delta \mathcal{L} \supset
  -\frac{\chi}{4}\,F'_{\mu\nu}F^{\mu\nu}
  -\frac{1}{4}\,F'^{\mu\nu}F'_{\mu\nu}
  -\frac{m_{\gamma'}^{2}}{2}\,A'_\mu A'^{\mu}.
\eeq
If the dark photon mass $m_{\gamma'}$ and the mixing parameter $\chi$ are non-zero with $|\chi|\ll 1$, the dark photon mixes with the ordinary photon.  
This mixing leads to an interesting photon-dark photon oscillation that can be exploited in various search strategies.

A representative setup is the light-shining-through-a-wall (LSW) experiment~\cite{Abel:2006qt,Adler:2008gk,Arias:2010bh,Redondo:2010dp,Bahre:2013ywa,Ortiz:2020tgs,Seong:2023ran,Hoof:2024gfk}.  
In this configuration, a light source is placed a distance $L_{1}$ from an opaque wall, while a photon detector is located a distance $L_{2}$ behind the wall.  
Photon-like states cannot penetrate the wall, but dark photon-like states can traverse it and reconvert into photons, yielding a detectable signal (see Fig.\,\ref{fig:setup}).  
In a one-dimensional treatment, the transmission probability for a single incident photon is conventionally estimated as
\beq
\laq{oscnom}
P_{\gamma\to\gamma}=16\,\chi^{4}\!\left(
  \sin\frac{\Delta k\,L_{1}}{2}\;
  \sin\frac{\Delta k\,L_{2}}{2}
\right)^{2},
\eeq
where
$\Delta k = E_\gamma-\sqrt{E_\gamma^{2}-m_{\gamma'}^{2}}
           \simeq m_{\gamma'}^{2}/(2E_\gamma)$
with $E_\gamma$ the photon energy.

Many experiments have already set limits using this technique~\cite{Redondo:2010dp,Ehret:2010mh,Inada:2013tx,Betz:2013dza,Halliday:2024lca};  
see also the LSW searches for axion-like particles (ALPs)~\cite{Battesti:2010dm,DellaValle:2015xxa,Inada:2016jzh,OSQAR:2015qdv}.  
Typically, a conversion region, such as a vacuum cavity or magnet bore, is provided immediately after the light source to enhance oscillations~\cite{Sikivie:2007qm}.  
Different light sources probe different regions of parameter space: optical lasers favor lower masses, whereas undulators, which generate X-rays, extend sensitivity to higher masses.

Indeed, a dark photon search using an undulator beamline at SPring-8 was performed in
Ref.~\cite{Inada:2013tx}, providing some of the strongest laboratory constraints on
comparatively heavy dark photons (see also the undulator-based ALP searches in
Refs.~\cite{Battesti:2010dm,Inada:2016jzh}).  
In this paper we likewise focus on the undulator as the light source, but instead
of occupying an experimental hutch, we consider placing a photon detector outside
the experimental hutch.

Recently, Ref.~\cite{Yin:2024rjb} noted that an undulator can also serve as a
source of ALPs because the strong internal magnets used to bend
the electron beam provide the $B$-field needed for ALP-photon conversion.  
A quantum-field-theoretic (QFT) analysis shows that adding a magnet and a photon
detector outside of the shield in front of the undulator automatically realizes an LSW configuration. With the production amplitudes of photons and any
beyond-standard-model (BSM) particles in hand, one can therefore perform a
complete study following \cite{Yin:2024rjb} without relying solely on the oscillation formula with monochromatic photon in one-dimension.

Indeed, sometimes by using the formula \eq{oscnom}, 
it is tacitly assumed that the relevant mass range of dark photon is limited to
$E_\gamma \gtrsim m_{\gamma'}$, so that $\Delta k$ remains real.
In realistic situations this assumption can fail at larger $m_{\gamma'}$, and
several caveats must be taken into account:
\begin{enumerate}
%[{\bf 1.}]
  \item \textbf{Wave-packet averaging} ---  
        The production amplitude must be integrated over the full momentum and
        angular distribution of the wave packet.  
        For sufficiently heavy
        $m_{\gamma'}$ the interference term is washed out, and the analysis
        should switch to the particle picture of mass eigenstates, as in
        accelerator experiments.
  \item \textbf{Source kinematics} ---  
        An undulator has an intrinsic microscopic structure and may fail to produce
        the heavy mode in the particle picture because of kinematic suppression.
  \item \textbf{Medium effects} ---  
        The wall (shield) has a material-dependent attenuation length, so the
        effective “shielding’’ is a non-trivial function of $m_{\gamma'}$ even if the energy is fixed. In addition, if we perform the experiment in air, one needs to consider the matter effect for the propagation. 
  \item \textbf{Detector response} ---  
        The photon detector itself has a characteristic structure and size and may not
        detect the dark photons na\"{i}vely via mixing.
\end{enumerate}
A precise assessment of these effects requires a concrete geometry and a careful
quantum-mechanical treatment; this is crucial for setting reliable sensitivity
limits and for interpreting past and future LSW results.

The purposes and main results of this paper are \textbf{twofold}:
\begin{itemize}
\item[{\bf I}.]  
  Using the QFT approach developed in \cite{Yin:2024rjb} together with the quantum-kinetic equation, we analyze dark photon and photon production in an undulator--wall(--air) configuration.  
  This treatment explicitly addresses caveats~{\bf 1--3} and shows that the resulting sensitivities differ significantly from the na\"{i}ve oscillation estimate.

\item[{\bf II}.]  
  Placing a photon detector along the beamline \emph{outside} the shielding provides excellent sensitivity, enabling an economical, parasitic LSW experiment that could set the strongest laboratory limits in the relevant mass range.  
\end{itemize}

This paper is organized as follows. In Sec.~\ref{chap:dark photon} I present the QFT-based formalism for photon and dark photon production in an undulator. In Sec.~\ref{chap:vac} I construct the full quantum‐mechanical treatment of a LSW setup, carefully including finite wave-packet effects, source kinematics, and material attenuation. In Sec.~\ref{chap:air} I extend the analysis to include air-filled propagation on the detector side, demonstrating the quantum‐Zeno suppression and resonance enhancement in medium. Finally, Sec.~\ref{chap:conclusion} is for the conclusions and discussion.

Throughout this paper I adopt the metric $\eta_{\mu\nu}\equiv\diag(1,-1,-1,-1)$ with $\mu,\nu=0,x,y,z$.  
Four-vectors are written as $V^{\mu}\equiv (V^{0},\vec{V})=(V^{0},V^{x},V^{y},V^{z})$.  
Three-vectors are $\vec{V}=(V^{x},V^{y},V^{z})$ and the norm is given by $V=|\vec{V}|$.
Natural units are used, in particular the speed of light is set to $c=1$.

\section{QFT for Dark-Photon and Photon Production from an Undulator}

\lac{dark photon}
An undulator (or wiggler) is a specialized photon source employed at synchrotron-radiation facilities to generate highly collimated, high-brightness light spanning the infrared to hard-X-ray and even $\gamma$-ray regimes. Relativistic electrons traverse a spatially periodic magnetic array, execute sinusoidal trajectories, and emit synchrotron radiation with characteristic wavelengths (see Fig.~1 of Ref.~\cite{Yin:2024rjb}).

Undulators at facilities such as the ESRF and SPring-8 have already been exploited in LSW experiments~\cite{Battesti:2010dm,Inada:2013tx,Inada:2016jzh}, providing more stringent constraints on heavy ALPs than searches based on optical lasers.

These measurements are, however, restricted by the modest length (a few meters) of the beamline hutches and by limited beam-time allocations, which preclude the installation of large-volume detectors and prohibit extended baselines. 
In addition, the photons in the experimental hutches are much smaller than the number of the primary photons directly produced by the undulator. 

In this paper I explore an alternative configuration. In the remainder of this section I estimate the production rate and wave functions of photons and dark photons emitted by the undulator.

\lac{motion}

\subsection{Electron Current in an Undulator}
The electron motion in an undulator can be modeled as follows:
\begin{itemize}
  \item The magnetic field in the region $z\in[0,L]$ is taken as
    \beq\laq{Bconfigu}
      \vec B_{\rm ext}(z)
      = B_{0}\,\bigl\{\cos(k z),\,\k\sin(k z+\phi),\,0\bigr\}\,,
    \eeq
    and vanishes for $z<0$ or $z>L$.  Here $k$ is the undulator wave number, $\k$ represent the ellipticity and
    $\phi$ the phase (e.g.\ for a helical undulator $\k=1,\;\phi=0$, while for a
    linear undulator $\k=0$).
  \item Solving the electron equation of motion yields
    \begin{align}
      \vec\beta
      &\simeq
      \Bigl(-\tfrac{K\,\k}{\gamma}\cos(k\beta^{z}t_{e}+\phi),
            -\tfrac{K}{\gamma}\sin(k\beta^{z}t_{e}),
            \beta^{z}\Bigr)\,,\nonumber\\
      \AND \label{sol}
      \vec r_{e}
      &\simeq
      \Bigl(-\tfrac{K\,\k}{\gamma\,k\,\beta^{z}}
            \sin(k\beta^{z}t_{e}+\phi),
            \tfrac{K}{\gamma\,k\,\beta^{z}}
            \cos(k\beta^{z}t_{e}),
            \beta^{z}\,t_{e}\Bigr)\,,
    \end{align}
    where $K\equiv  \frac{B_0e}{k m_e}$ is the so-called K-parameter. When $K\lesssim1$, the light source is called a undulator (if $K\gg1$ it is called a wiggler).
  \item The longitudinal velocity, $\beta^z$, and Lorentz factor, $\g$, satisfy
    \beq
      \beta\simeq\beta^{z}\simeq\sqrt{1-\gamma^{-2}}\,.
    \eeq
\end{itemize}

\subsection{Wave function of undulator (dark) photons}
\lac{DP}

We are now ready to estimate dark photon and photon productions.  
Throughout this section I treat the produced particles as asymptotic states.  
I assume that the undulator is embedded in an infinitely large volume so that the particle momenta—and hence their masses—can in principle be measured with arbitrary precision.  
Consequently, I work in the mass‐eigenstate basis when analyzing particle production.

I mainly concentrate on the production of the heavier mass eigenstate $\ket{2}$ (approximately the dark photon) and comment on the production of the lighter state $\ket{1}$ (approximately the photon).

To estimate heavier state production I switch to the kinetically normalized basis and couple the mass eigen-field, $A'_\m$, to a classical current
\beq
\laq{Lag}
{\cal L}_{\rm int}=-\chi\,A'^\mu j_\mu,
\qquad
j_\mu=-e\,v_\mu(t)\,\delta^{3}\!\bigl(\vec x-\vec r_{e}[t]\bigr),
\eeq
where $v^\mu=(1,\vec\b)$ is the electron four-velocity.

The production amplitude is
\beq
\laq{amp}
\langle 2,\e,k_{\g'}|0_{j}\rangle
=-\int d^{4}x\,
\frac{i\,\e^\mu}{\sqrt{2w_{\g'}}}\,
j_\mu\,
e^{i\bigl(w_{\g'} t-\vec k_{\g'}\!\cdot\!\vec x\bigr)}
=-\frac{i}{\sqrt{2w_{\g'}}}\,
\e^\mu\,\tl j_\mu(k_{\g'}),
\eeq
where $\e^\mu$ is the polarization vector,  
$\vec k_{\g'}$ is the dark photon momentum,  
and $w_{\g'}=\sqrt{k_{\g'}^{2}+m_{\g'}^{2}}$ is its energy.  
The state $\ket{0_{j}}$ denotes the vacuum in the presence of the external current~$j$.  
The Fourier transform of the current is defined as
\beq
\laq{tljm}
\tl j_\mu(q)\equiv
-e \int_{-\infty}^{\infty}\!dt\;
v_\mu(t)\,
e^{i\bigl(q^{0}t-\vec q\!\cdot\!\vec r_{e}[t]\bigr)}.
\eeq

Neglecting ${\cal O}(K^{2},\g^{-2})$ terms so that charge conservation $q^\mu\tl j_\mu(q)=0$ is explicit, I obtain~\cite{Yin:2024rjb}
\begin{align}
\frac{\tl j^{0}(q)}{\pi}&=-e\!\Bigl[
  2\delta(\Delta q)
 +\bigl(q^{x}\k e^{i\f}-i q^{y}\bigr)\frac{K}{2\g k}\,
   \delta_{\frac{L}{\b^{z}}}(\Delta q+k\b^{z})
 -\bigl(q^{x}\k e^{-i\f}+i q^{y}\bigr)\frac{K}{2\g k}\,
   \delta_{\frac{L}{\b^{z}}}(\Delta q-k\b^{z})
 \Bigr],\\[4pt]
\frac{\tl j^{x}(q)}{\pi}&=\frac{eK{\b}}{2\g}\!
 \Bigl[
   \k e^{i\f}\,\delta_{\frac{L}{\b^{z}}}(\Delta q+\b^{z}k)
  +\k e^{-i\f}\,\delta_{\frac{L}{\b^{z}}}(\Delta q-\b^{z}k)
 \Bigr],\\[4pt]
\frac{\tl j^{y}(q)}{\pi}&=\frac{eK{\b}}{2\g i}\!
 \Bigl[
   \delta_{\frac{L}{\b^{z}}}(\Delta q+\b^{z}k)
  -\delta_{\frac{L}{\b^{z}}}(\Delta q-\b^{z}k)
 \Bigr],\\[4pt]
\frac{\tl j^{z}(q)}{\pi}&=-e\b^{z}\!\Bigl[
  2\delta(\Delta q)
 +\bigl(q^{x}\k e^{i\f}-i q^{y}\bigr)\frac{K}{2\g k}\,
   \delta_{\frac{L}{\b^{z}}}(\Delta q+k\b^{z})
 -\bigl(q^{x}\k e^{-i\f}+i q^{y}\bigr)\frac{K}{2\g k}\,
   \delta_{\frac{L}{\b^{z}}}(\Delta q-k\b^{z})
 \Bigr],
\end{align}
with
\beq
\Delta q \equiv q^{0}-\b^{z}q^{z},
\eeq
and
\beq
\laq{delta}
\pi\delta_{X}(q)\equiv\int_{0}^{X}\!dx\,e^{iqx}
=\frac{e^{iqX}-1}{iq},
\eeq
which acts as a finite-width delta function.

In the numerical simulations, I introduce an exponential regulator,
\beq
\pi\delta_{L}(q)=\frac{e^{iqL}-1}{q}\;\longrightarrow\;
\frac{e^{iqL}-1}{q}\,e^{-\frac{\epsilon\,\ab{q}L}{2\pi}},
\eeq
to suppress the unphysical enhancement that appears at $k_{\g'}\gg\g^{2}k$ and $\theta\sim1$.  
I have confirmed that a comparable regularization is obtained by exponentially damping the electron’s transverse motion near the boundaries $z\simeq0,L$.  
Without such a regulator, the abrupt change in the electron trajectory excites spurious high-momentum modes in Fourier space.\footnote{These effects should also be taken into account when studying the high-mass tail of generic LSW experiments, as they modify the boundary behavior at large masses.}  
Throughout the simulations I set $\epsilon=1/20$.

Then, \Eq{amp} becomes
\begin{align}
\langle 2,\epsilon,k_{\gamma'} | 0_j \rangle
&\simeq i\,\frac{e\,\chi\,K\pi}{\sqrt{2w_{\gamma'}}\,2k\gamma}\!
\left[
      k\!\left(\kappa e^{-i\phi}\epsilon^{x}+i\epsilon^{y}\right)
     +\epsilon^{z}\!\left(\kappa e^{-i\phi}k_{\gamma'}^{x}+ik_{\gamma'}^{y}\right)\beta^{z}
\right]
\,
\delta_{\frac{L}{\beta^{z}}}\!\bigl(\Delta k_{\gamma'}-k\beta^{z}\bigr)
\laq{ampsimp}
\end{align}
Here $\cos\theta\equiv k_{\gamma'}^{z}/k_{\gamma'}$.  
In deriving the amplitude I have discarded the terms proportional to
$\delta(\Delta k_{\gamma'})$ and
$\delta_{\frac{L}{\beta^{z}}}(\Delta k_{\gamma'}+k\beta^{z})$,
because their arguments never vanish in the regime of interest.  

As is customary, I set $\epsilon^{0}=0$.  
The two transverse polarization vectors are  
\beq
\vec\epsilon_{\mp}=\frac{1}{\sqrt{2}}
\bigl(
 -\cos\theta\cos\tilde\phi,\,
 -\cos\theta\sin\tilde\phi,\,
  \sin\theta
\bigr)
\;\mp\;
\frac{i}{\sqrt{2}}
\bigl(
  \sin\tilde\phi,\,
 -\cos\tilde\phi,\,
  0
\bigr),
\eeq
where $\tilde\phi$ satisfies  
$k_{\gamma'}^{x}=k_{\gamma'}\sin\theta\cos\tilde\phi,\;
 k_{\gamma'}^{y}=k_{\gamma'}\sin\theta\sin\tilde\phi,\;
 k_{\gamma'}^{z}=k_{\gamma'}\cos\theta$.  
The longitudinal polarization is  
$\epsilon^{\mu}_{\rm L}=\bigl(k_{\gamma'}/m_{\gamma'},
 \,w_{\gamma'}\vec k_{\gamma'}/(m_{\gamma'}k_{\gamma'})\bigr)$,  
but it is irrelevant here because its contribution is suppressed by
the small mass:  
$\epsilon^{\mu}_{\rm L}(k_{\gamma'})=k_{\gamma'}^{\mu}/m_{\gamma'}+
{\cal O}(m_{\gamma'}/k_{\gamma'})$ and
$k^{\mu}_{{\g'}}\tl j_{\mu}(k_{\g'})=0$ for an on-shell electron according to Ward-Takahashi identity. 

The dark photon spectrum is narrowly peaked around
\beq
\laq{kg}
k_{\gamma'}\simeq\frac{2k}{\gamma^{-2}+\theta^{2}}-\frac{m_{\gamma'}^{2}}{2k},
\eeq
obtained from $\Delta k_{\gamma'}-k\beta^{z}=0$ expanded in
$\gamma^{-1}$, $\theta$, and $m_{\gamma'}$.  
It is not strictly monochromatic; its line shape follows
$\delta_{L/\beta^{z}}(\Delta k_{\gamma'}-k\beta^{z})$,  
whose finite width $1/L$ translates into an energy spread
$\Delta k_{\gamma'}\sim(kL)^{-1}k_{\gamma'}$.%
\footnote{Electron-beam energy spread and the $K$-parameter also broaden the line, but for moderate $L$ the dominant effect comes from the finite undulator length.}  
Accordingly, throughout this paper I retain the finite-width form
$\delta_{L/\beta^{z}}(\Delta k_{\gamma'}-k\beta^{z})$ instead of
replacing it by a Dirac delta, unlike Ref.~\cite{Yin:2024rjb}.

Because I consider only transverse modes, the same expression applies
to ordinary photons in the limit $m_{\gamma'}\!\to0$ and $\chi\!\to1$:\footnote{This is the case when the leading order approximation of $\chi$ in $\bigl|\langle 2,\epsilon,\vec k_{\gamma'}|0_j\rangle\bigr|^{2}$ is considered. With higher order of $\chi$ for the mixing, this limit does not give the relation.}
\beq
\bigl|\langle 1,\epsilon,\vec k_{\gamma'}|0_j\rangle\bigr|^{2}
=\bigl|\langle 2,\epsilon,\vec k_{\gamma'}|0_j\rangle\bigr|^{2}_
{\chi\to1,\;m_{\gamma'}\to0}. \laq{photonlimit}
\eeq

\subsection{Features of undulator (dark) photon spectra}
\lac{spectra}

\begin{figure}[t!]
  \bec
    \includegraphics[width=75mm]{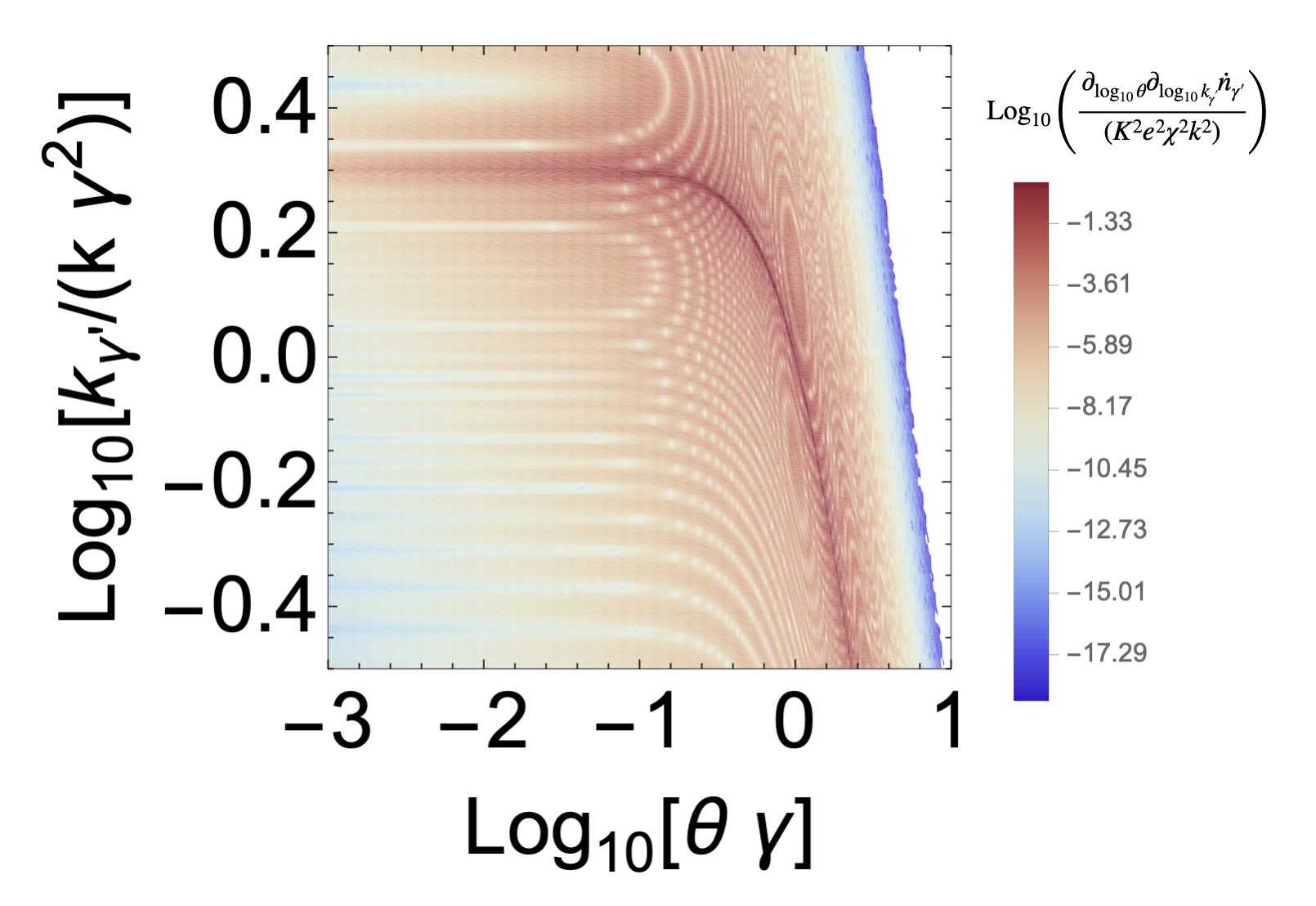}
    \includegraphics[width=70mm]{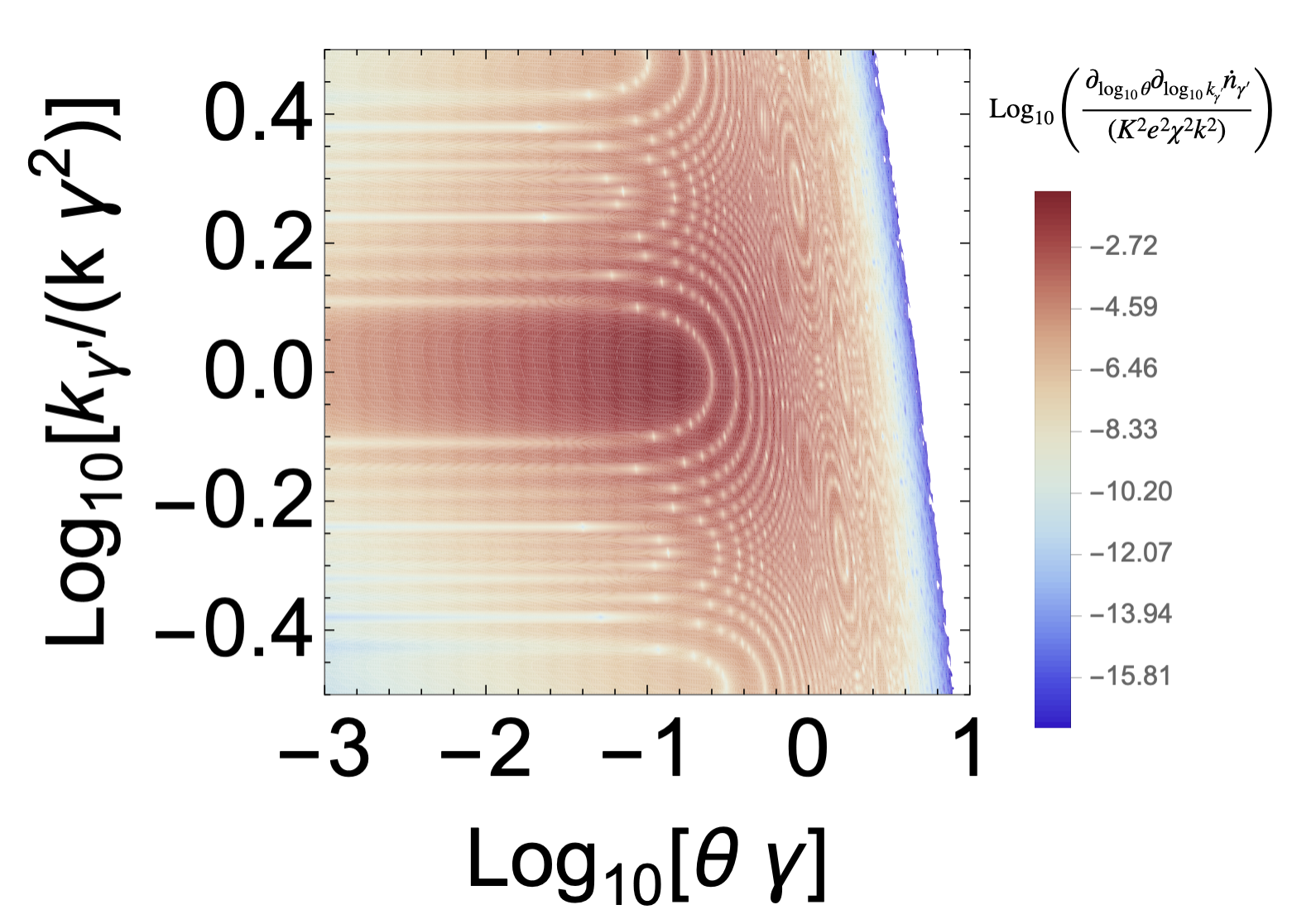}
  \eec
  \caption{Momentum and polar--angle distributions of undulator (dark) photons.
  The left panel corresponds to $m_{\g'}=0$, while the right panel shows the case $m_{\g'}=k\g$.
  Here $\g=6\times10^{3}$ and $L=200/k$.}
  \label{fig:1}
\end{figure}

\begin{figure}[t!]
  \bec
    \includegraphics[width=105mm]{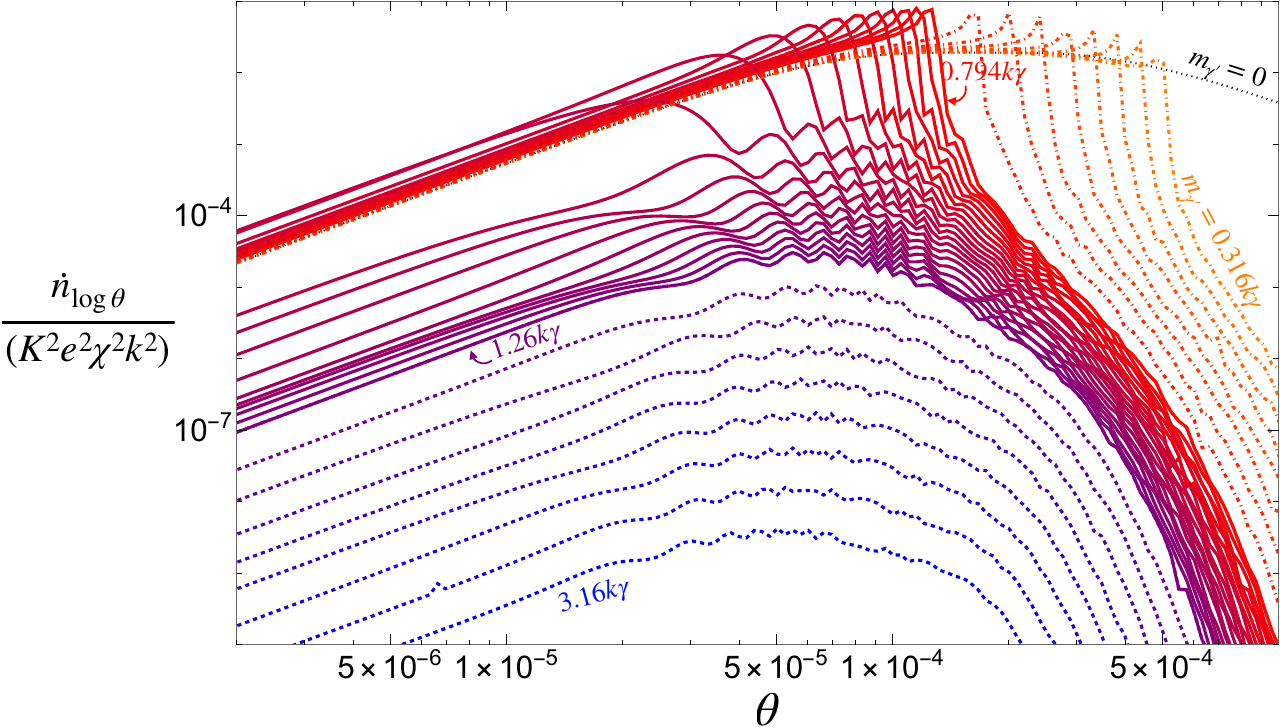}
  \eec
  \caption{$\dot{n}_{\log\theta}$ by varying $\theta$ for various $m_{\g'}$.
  The dotted black line denotes $m_{\g'}=0$.
  Dot--dashed colored lines correspond to $m_{\g'}/k\in[0.316,0.794)$
  (equal logarithmic spacing); solid lines to $m_{\g'}/k\in[0.794,1.26]$;
  dashed lines to $m_{\g'}/k\in(1.26,3.16]$.
  In all cases $\g=6\times10^{3}$ and $L=200/k$.}
  \label{fig:3}
\end{figure}

\begin{figure}[t!]
  \bec
    \includegraphics[width=105mm]{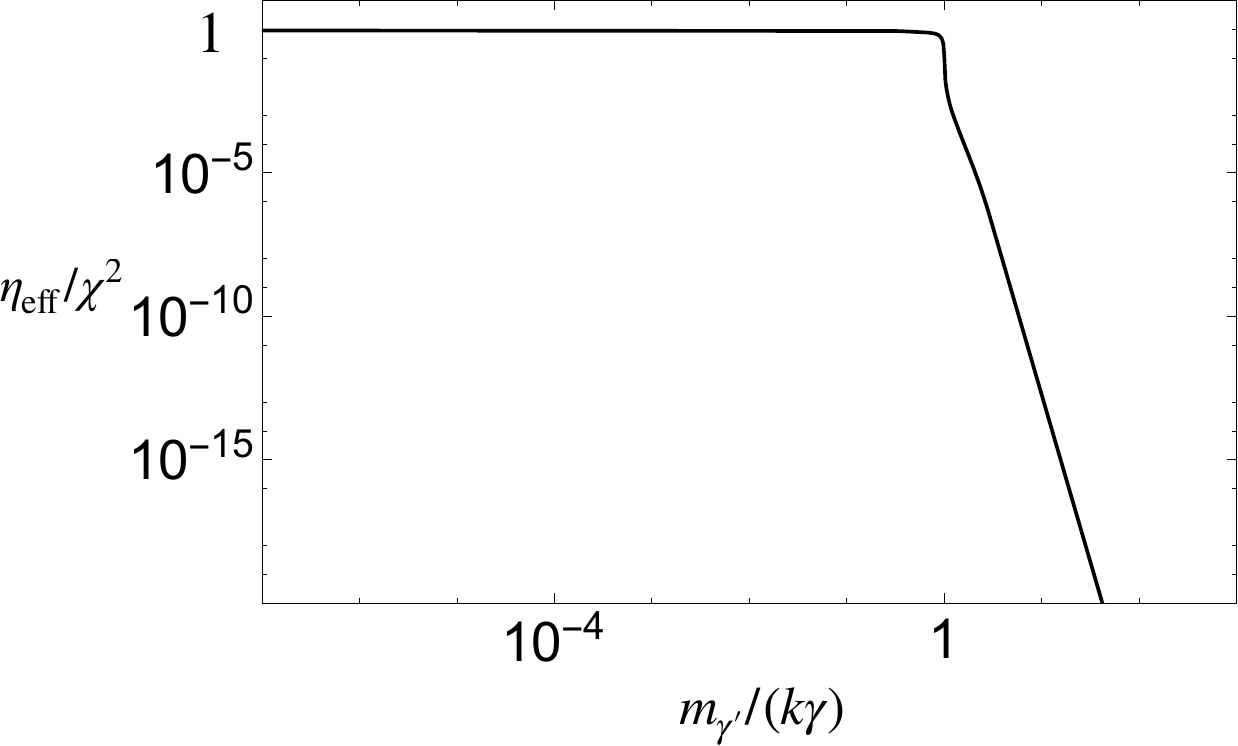}
  \eec
  \caption{Effective photon-dark photon conversion efficiency for the massive mode production as a function of $m_{\g'}$. I take $\g=6\times10^{3}$ and $L=200/k$.}
  \label{fig:4}
\end{figure}

The differential production rate for the transverse modes of the massive state is  
\beq
\laq{flux}
\partial_{\Omega}\dot n_{2}
  =\frac{\b}{L}
   \int\!\frac{dk_{\g'}\,k_{\g'}^{2}}{(2\pi)^{3}}
   \sum_{\e^{\pm}}
   \ab{\langle 2,\e,\vec k_{\g'}|0_{j}\rangle}^{2},
\eeq
where \Eq{ampsimp} is used for the amplitude.
Keeping only the leading terms in the $K^{2}$ and $1/\g$ expansions, the rate factorizes as
\beq
\partial_{\Omega}\dot n_{2}
  \;\approx\;
  \chi^{2}\,\F(\tl\f)\;
  \times\;
  \frac{1}{2\pi}\,\dot n_{\theta},
\eeq
with
\beq
\F(\tl\f)=\ab{\k\cos\tl\f-i\,e^{-i\f}\sin\tl\f}^{2}.
\eeq
For convenience I also define
\beq
\dot n_{\log_{10}\theta}\equiv\log(10)\,\theta\,\dot n_{\theta},
\qquad
\partial_{\log_{10}k_{\g'}}\dot n_{\log_{10}\theta},
\eeq
the latter being the spectrum before integrating over $k_{\g'}$.

For a helical undulator, $\k=1$ and $\f=0$, so $\F(\tl\f)=1$ and therefore  
\beq
\partial_{\Omega}\dot n_{2}
  =\chi^{2}\,\frac{\dot n_{\theta}}{2\pi}.
\eeq
The following discussion assumes this helical configuration; results for other
undulator types can be recovered by reinstating the factor $\F(\tl\f)$.\footnote{In the following sections, the derived sensitivities do not depend on $\k $ and $\f$. For instance, in \Eq{etaeff} the coefficient from the $\tl \f$ integral cancels out. However it can be important for alleviating the background if the detector has a good angular resolution.  }

The massless state flux (approximately the photon) is obtained from \Eqs{flux} and \eq{photonlimit} by taking the massless limit and setting $\chi=1$:
\beq
\partial_{\Omega}\dot n_{1}
  \;\approx\;
  \partial_{\Omega}\dot n_{\g}
\approx
  \left.
    \partial_{\Omega}\dot n_{2}
   \right|_{m_{\g'}\to0,\;\chi=1}.
\eeq

\paragraph{(Dark) photon spectra}

We now examine the spectral features in detail.  For numerical illustration I set $\gamma=6\times10^{3}$ and $L=200/k$, values representative of state-of-the-art synchrotron undulators.  Fig.~\ref{fig:1} shows the integrand of \Eq{flux}, $\partial_{\log_{10}k_{\g'}}\dot n_{\log_{10}\theta}$, for two benchmark masses.  The left panel corresponds to the massless limit $m_{\g'}=0$ (equally valid for dark photons with $m_{\g'}\!\ll\!k$ and $K\lesssim1,\chi=1$), whereas the right panel depicts the heavy case $m_{\g'}=k\gamma$.  When $m_{\g'}$ approaches $k\gamma$, so that the oscillation frequency is $\sim(k\gamma)^{2}/(k\gamma^{2})\simeq k$, the spectrum changes qualitatively and the production rate itself is modified.  This shows clearly that production can no longer be described by the na\"ive photon-dark photon oscillation formula \Eq{oscnom}. 
  In particular, the momentum of the dark photon at $\theta \sim 0$ is no longer $2k\gamma^{2}$ (the photon value) but is shifted to lower values.

This behavior is clarified in the electron  comoving frame, where the longitudinal electron momentum vanishes.  There the electron oscillates with phase $\beta k\gamma t'$, 
with $t'$ being the time in the comoving frame, 
 so the radiated frequency is $\sim k\gamma$, not $k\gamma^{2}$.  Consequently, production of dark photons with $m_{\g'}\gtrsim k\gamma$ is suppressed by energy-momentum conservation; such states arise only through the violation of energy conservation due to the finite interaction time as the magnets pass the electron.  Moreover, dark photons with $m_{\g'}\sim k\gamma$ are produced non-relativistically, giving a laboratory energy of $k\gamma^{2}$ rather than $2k\gamma^{2}$.

For ordinary photons (left panel of Fig.~\ref{fig:1}) the emitted energy depends on the polar angle $\theta$.  Because the solid-angle integral is dominated by $\theta\sim1/\gamma$, the typical photon energy contributing to the flux is $k\gamma^{2}$, not the $2k\gamma^{2}$ value attained in the $\theta\!\to\!0$ limit, which is sometimes adopted in literatures for the synchrotron radiation, aiming the use of the monochromatic photons with $\theta\to 0$.

\paragraph{Effective photon-dark photon conversion rate}

Integrating $\partial_{\log_{10}k_{\g'}}\dot n_{\log_{10}\theta}$ over $\log_{10}k_{\g'}$ yields $\dot n_{\theta}$ for each mass.  Fig.~\ref{fig:3} plots $\dot n_{\log\theta}$ as a function of $\theta$.  The black dotted curve is the massless case $m_{\gamma'}=0$.  Colored dot-dashed lines correspond to $m_{\g'}/k\in[0.316,0.794)$, solid lines to $m_{\g'}/k\in[0.794,1.26]$, and dashed lines to $m_{\g'}/k\in(1.26,3.16]$, each set shown with equal logarithmic spacing.  A pronounced enhancement appears at $\theta\ll1/\gamma$ when $m_{\g'}\simeq k\gamma$, which is analytically discussed in Appendix~\ref{app:circular}.

However, the enhancent disappears once we integrate over momentum.  
This can be seen by plotting
\beq
\laq{etaeff}
\eta_{\rm eff}(m_{\g'})\;\equiv\;
\frac{\dot n_{\g'}(m_{\g'})}{\dot n_{\g}}
= \frac{\dot n_{\g'}(m_{\g'})}{\left.\dot n_{\g'}(0)\right|_{\chi=1}},
\eeq
as a function of $m_{\g'}$; see Fig.~\ref{fig:4}.%
\footnote{Hence the enhancement due to the mass effect is not important for the total signal when the detector is large enough to cover the full angular spread, which is assumed in this paper.  It can, however, be relevant for detectors with fine angular/energy resolution or limited acceptance.}

As we shall show below, the production of the massless mode is just as important as the massive mode for determining the LSW sensitivity.

\section{Quantum study of an LSW experiment for dark photon}

We now present a detailed quantum-mechanical analysis of the photon-dark photon system in an LSW configuration, aiming to derive sensitivity limits without neglecting various scale-dependent effects that are often omitted.

\begin{figure}[t!]
  \begin{center}  
\includegraphics[width=145mm]{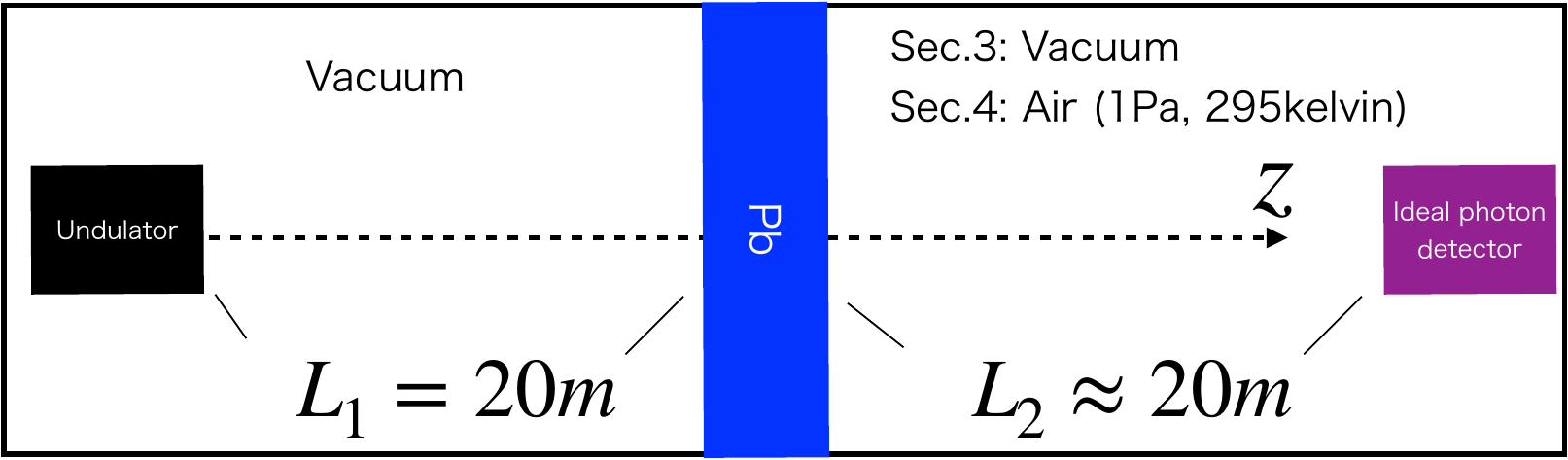}
  \end{center}
    \caption{The systematical view of the setup of consideration. 
I consider   $L_1\approx L_2\approx 20$m and $\k=1, \f=0,K\ll1, k=2\pi/(0.03{\rm m}), L=200/k\approx 1\rm m$ and $\g=6000.$ 
    In \Sec{vac} I consider the detector side is in the vacuum. 
    In \Sec{air}, the detector side is filled with air. In this case the result does not depend on the size of $L_2$ if it is larger than the $1/\mu_{\rm air}.$}
    \label{fig:setup}
    \end{figure}

For clarity, I first consider the idealized case in which both the undulator and the photon detector operate in vacuum, so matter effects from air can be neglected.  (\Sec{air} treats the more realistic configuration with air on the detector side.)  A schematic of the arrangement is shown in Fig.~\ref{fig:setup}. 

Throughout this section I adopt
$L_1\approx L_2\approx 20$m and $\k=1, \f=0,K\ll1, k=2\pi/(0.03{\rm m}), L=200/k\approx 1\rm m$ and $\g=6000.$ 
so the typical photon energy is $k_{\g}\sim\mathrm{keV}$, representative of  many beamlines in synchrotron facilities.

We assume the wall is made of lead.  The linear attenuation coefficient is then
\[
\mu_{\rm pb}\,\simeq\,2\times10^{4}\;\mathrm{cm}^{-1}
\]
in the keV range~\cite{lemmon2010nist}.  The corresponding plasma mass is $m_{\g,\rm p,pb}^{2}$. The resulting oscillation length $2k_{\g'}/m_{\g,\rm p,pb}^{2}$ are of order $1/\mu_{\rm pb}$.  Both $\mu_{\rm pb}$ and $m_{\g,\rm p,pb}^{2}$ can be extracted from the real and imaginary parts of the complex refractive index (see Fig.~\ref{fig:muair} and \Eq{reflectiveindex}).

I assume a photon detector that registers the massless and massive eigenstates with probabilities
\beq
P_{\rm det}=1\AND\chi^{2},
\eeq
respectively.  In other words, detection is governed purely by the photon‐flavor component and is taken to be independent of the heavier eigenstate’s mass.\footnote{This assumption is reasonable for many detector technologies.  For example, in silicon detectors the signal is the number of electrons liberated by electron-hole pair creation; the threshold for such pair creation is typically ${\cal O}(\EV)$, well above the scale $\mu_{\rm pb}\sim0.1\EV$ at which our sensitivity will drop.  Detectors based on atomic excitation have even higher thresholds, ${\cal O}(10)\EV$.
Treating $P_{\rm det}=\chi^{2}$ for the heavier eigenstate is even conservative in some cases (see \Sec{conclusion}). 
} I will also assume that the detector covers the region transverse to the beamline with a radius much larger than $10(L_{1}+L_{2})/\g \approx 0.07$m, so all relevant emission angles are collected.  

A detailed background study is deferred to future work, as it depends on the specific detector technology and experimental environment.  As a reference, Ref.~\cite{Inada:2013tx} reports a background rate of $\O(10^{4})/{\rm yr}$ in the $5$-$30\KEV$ range (bin width $\O(0.1)\KEV$) inside the SPring-8 experimental hutch.  Assuming a comparable background for a year-long run, a discovery would require a signal of $\O(10^2-10^3)/{\rm yr}$.  Accordingly, I adopt two benchmark signal rates: $1/{\rm yr}$ (optimistic) and $10^{4}/{\rm yr}$ (conservative).

\lac{vac}

\subsection{Summary of the result}
\begin{figure}[t!]
  \begin{center}  
\includegraphics[width=145mm]{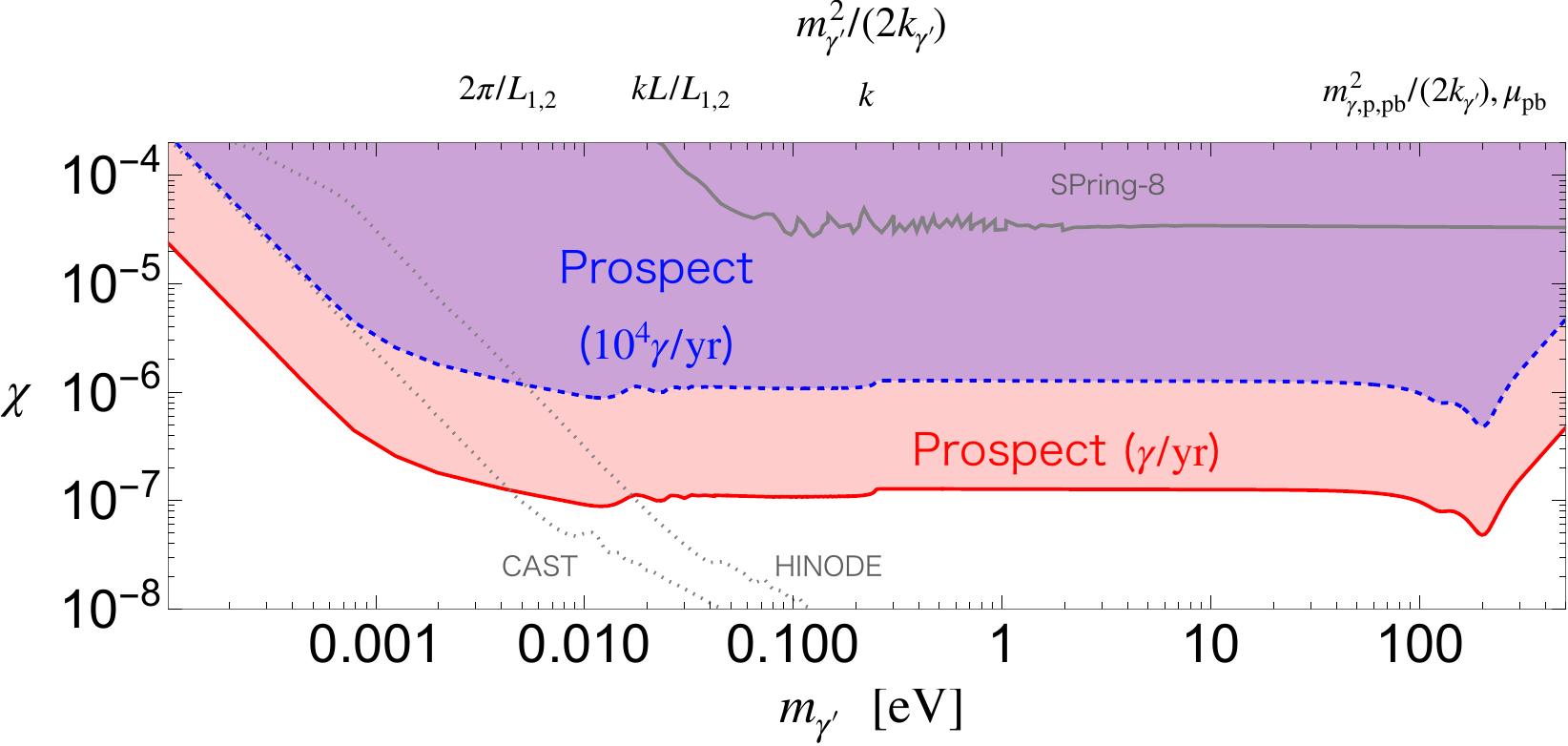}
  \end{center}
    \caption{Prospects of the LSW search of dark photon from undulator. 
 $\g=6\times 10^3, L=200/k, 2\pi/k=0.03$m, $\dot{N}_{\g}\approx 2\times 10^{27}/\rm yr$ are assumed.  $L_1\approx L_2\approx 20$m is assumed. 
 The top frame denotes the relevant scales in terms of flavor oscillation frequency. 
The existing laboratory limit is taken from Ref.\,\cite{Inada:2013tx}. The reach can be also competitive with some astrophysical limits~\cite{Redondo:2008aa, Frerick:2022mjg}. }
    \label{fig:limit}
    \end{figure}

Before showing the calculation details given in the following subsections, here I show the main result. Fig.~\ref{fig:limit} displays the resulting sensitivity in the $(m_{\g'},\chi)$ plane.  The prospects for $1$ detected photon$/{\rm yr}$ ($10^{4}$ photons/yr) are shown by the red solid (blue dashed) curves.  For comparison, the existing LSW limit appears as the grey dot-dashed line in the top, while selected solar dark photon bounds are drawn with dot-dashed colors.  

The figure’s structure reflects the non-trivial effects discussed as caveats {\bf 1-3} in the Introduction and the projected reach differs markedly from the ones that can be derived from the na\"ive \Eq{oscnom}.  
To explain this, 
 let us express the hierarchy of the relevant scales in terms of the oscillation length $2k_{\g'}/m_{\g'}^{2}$
\beq
\frac{2\pi}{L_{1,2}}\;\ll\;\frac{kL}{L_{1,2}}\;\ll\;k
\;\ll\;
\frac{m_{\g,\rm p,pb}^{2}}{2k_{\g'}}, \mu_{\rm pb}\;(\ll k_{\g'}).
\eeq
The corresponding scale is also shown in the top panel in Fig.\ref{fig:limit}. 

Near $2\pi/L_{1,2}$ the onset of oscillations already deviates from the na\"{i}ve \Eq{oscnom} because the photon spectrum is not monochromatic and varies strongly with polar angle (cf. the left panel of Fig.~\ref{fig:1}); wave-packet smearing broadens the threshold.

The scale $kL/L_{1,2}$ marks decoherence within the wave packet (caveat~{\bf 1.}).  Decoherence occurs when
\beq
\laq{criterion}
\Delta k_{\g'}\;\partial_{k_{\g'}}\!\left(\frac{m_{\g'}^{2}L_{1,2}}{2k_{\g'}}\right)\gg2\pi,
\eeq
where $\Delta k_{\g'}\sim k_{\g'}/(Lk)$ is the packet width inferred from the finite-length factor in the ``delta function'' in \Eq{ampsimp}.  

$k$, corresponding to $m_{\g'}\sim k \g$, is the characteristic scale at which production of the massive mode is suppressed (see Fig.~\ref{fig:4}).  
Although the heavy eigenstate is impeded, massless states that impinge on the wall can regenerate the dark photon flavor component, contributing to the sensitivity even for $m_{\g'}\gtrsim k\gamma$.  
When this contribution is included, the overall suppression manifests as the slight kink in the sensitivity curve of Fig.\,\ref{fig:limit}.

For $\frac{m_{\g'}^{2}}{2k_{\g'}}\gtrsim\mu_{\rm pb}$ the sensitivity falls, and the loss sets in. This happens with $m_{\g'}\ll k \g^2$.  Conversely, when $m_{\g'}\simeq m_{\g,\rm p,pb}$ a resonance in the wall enhances the reach substantially.

Therefore, the result shows a very different structure than the usually adopted sensitivities using \Eq{oscnom}. 
More detail scale-dependent analysis and discussion will be shown in the following. 

%for the mass eigenstate \Eq{criterion}, by using $E_\g\sim k\g^2$ derived previously. 
%This is possible if the detector size, $d_L$, is small enough,  $m_{g'}^2/k_{\g'}\times d\ll 1,$ that the quantum mechanically the dark photon mass is negligible.   
%As I will show that the physics relevant to the scales significantly alter the photon-dark photon conversion, and the the experimental and theoretical study on the LSW for the dark photon must include those consideration.

\subsection{Quantum mechanics of photon-dark photon oscillation: $m_{\g'}^{2}/k_{\g'}\ll k$}
\lac{qosc}

Consider the mass range
\beq
m_{\g'}\ll k\gamma,
\eeq
for which the dark‐photon mass is negligible inside the undulator.  
In this regime it is convenient to work in the flavor basis.  
The \emph{photon flavor} state, $\ket{\g}$, couples directly to the electron and is orthogonal to the \emph{dark‐photon flavor} state, $\ket{\g'}$, which does not.

The photon flavor is produced with amplitude
\beq
\tl\F(\e,\g,\vec k_{\g'})
  \;\simeq\;
  \langle 1,\epsilon,k_{\g'}|0_{j}\rangle
  \;\simeq\;
  \langle 2,\epsilon,k_{\g'}|0_{j}\rangle\bigl/\chi^2.
\eeq
where I neglect the mass correction.  

The wave function of the dark‐photon flavor at time $t$ is then
\beq
\F_{\g'}(\e,\vec x,t)
  \;\equiv\;
  \sum_{\alpha=1,2}
  \int\!\frac{d^{3}\vec k_{\g'}}{(2\pi)^{3}(2w_{\alpha})}\,
  c_{\alpha}\,
  e^{i(\vec k_{\g'}\cdot\vec x-w_{\alpha}t)}\,
  \langle {\alpha},\epsilon,k_{\g'}|0_{j}\rangle,
\eeq
where $\alpha$ labels mass eigenstates and $c_{\alpha}$ are the corresponding coefficients $c_{1}\approx -\chi, c_2 \approx 1$. Again the mass correction in $\bra{\a,\e,k_{\g'}}0_j\rangle$ is neglected.
% for the ease of estimation. 

Only this $\ket{\g'}$ component can traverse the wall.  
Therefore I estimate the total flux through the $x,y$-plane at $z=L_{1}$
\begin{align}
N_{\g'}
  &=\sum_{\e^{\pm}}
    \int_{z=L_{1}}\!dt\,dx\,dy\,
    \F_{\g'}^{*}(\e,\vec x,t)\,
    (-i\partial_{z})\,
    \F_{\g'}(\e,\vec x,t)
    +\text{h.c.}\nonumber\\[4pt]
  &\approx
    \sum_{\e^{\pm}}\chi^{2}
    \int\!\frac{d^{3}\vec k_{\g'}}{(2\pi)^{3}}\,
    \ab{\tl\F(\e,\g,\vec k_{\g'})}^{2}
    \,
    4\sin^{2}\!\Bigl[
      \tfrac{1}{2}\bigl(k_{\g',z}-k_{\g,z}\bigr)L_{1}
    \Bigr],
\end{align}
where $k_{\g',z}=\sqrt{k_{\g'}^{2}-k_{\perp}^{2}-m_{\g'}^{2}}$ and $k_{\g,z}=\sqrt{k_{\g'}^{2}-k_{\perp}^{2}}$.  
The second line follows from performing the $t$, $x$, and $y$ integrals, which enforce energy and transverse‐momentum conservation.
% before integrating over $\vec k_{\g'}$.
%\section{Results} 

%Notice that in this region, only the propagation behavior affects the final result. 
The dark photon flavor just passing through the wall has the wave function  consistent with the flux of $\frac{d^2 P_{\g'}}{d \Omega d k_{\g'}}\approx \sum_{\e^{\pm}}\frac{1}{(2\pi)^3} \ab{\tl \F(\e,\g,\vec k_{\g'})}^2 
4\sin^2\(
\frac{m_{\g'}^2}{4k_{\g'}}
%\frac{\(k_{\g',z}-k_{\g,z}\)}{2}
L_1\)
$, where I took $m_{\g'}\ll k_{\g'}.$ Then I can repeat the previous estimation by replacing $\g\leftrightarrow \g'$ and considering $\tl{\F}(\e,\g',\vec{k}_{\g'})$ satisfying $|\tl{\F}(\e,\g',\vec{k}_{\g'})|^2= \ab{\tl \F(\e,\g,\vec k_{\g'})}^2 
4\sin^2\(
\frac{m_{\g'}^2}{4k_{\g'}}
%\frac{\(k_{\g',z}-k_{\g,z}\)}{2}
L_1\).
$ 
To include the conversion to photon flavor in the detector side at $z=L_1+L_2$, I get 
\beq
P^{\rm eff,light}_{\gamma\to\gamma}
\approx \frac{1}{L\dot{n}_{\g}}\chi^4 \sum_{\e^{\pm}}\int \frac{d^3 \vec k_{\g'}}{(2\pi)^3} \ab{\tl \F(\e,\g,\vec k_{\g'})}^2 
4\sin^2\(
\frac{m_{\g'}^2}{4k_{\g'}}
L_1\)\times
4\sin^2\(
\frac{m_{\g'}^2}{4k_{\g'}}
L_2\).\laq{lmass}
\eeq
Here I devide by $L\dot{n}_{\g}/\b^z \approx L\dot{n}_{\g}$ which is the number of photon that are emitted per each electron going through the undulator, and thus this is the effective transmission probability: the probability that when the detector receives the photon after a photon flavor is emitted from the undulator. 
This agrees with the oscillation formula \Eq{oscnom} averaged by using the wave packet. 
Note that the momentum is emitted mostly in the forward direction. 

\begin{figure}[t!]
  \begin{center}  
\includegraphics[width=105mm]{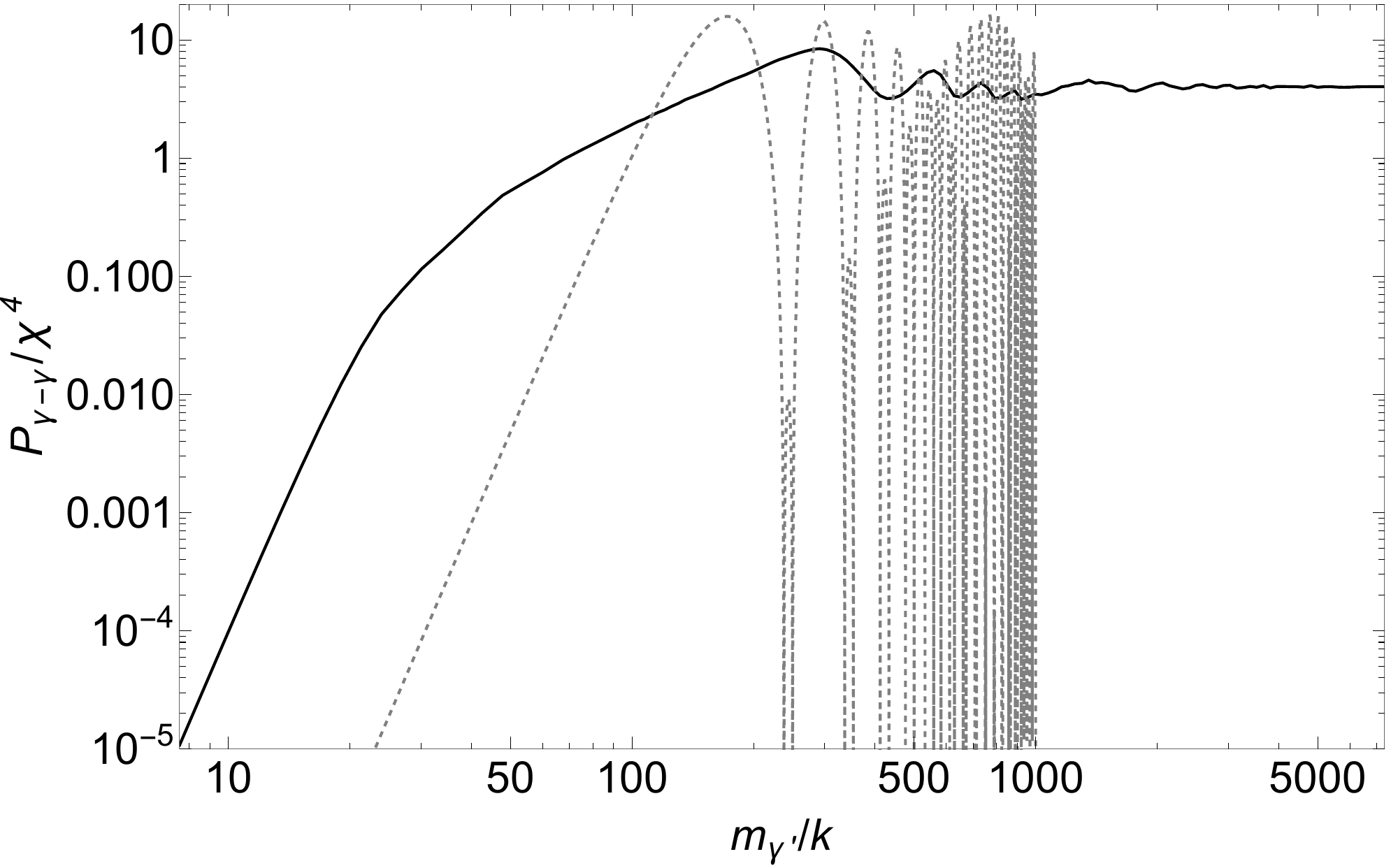}
  \end{center}
    \caption{$P^{\rm eff}_{\g\to \g}$ by varying $m_{\g'}$ for the photon produced from undulator. 
 $\g=6\times 10^3, L=200/k$ are taken. }
    \label{fig:osc}
    \end{figure}

$P^{\rm eff, light}_{\g\to\g}$ is plotted as the black solid curve in Fig.~\ref{fig:osc} as a function of $m_{\g'}$.  
Here I take $\gamma=6\times10^{3}$, $L=200/k$, $L_{1}=20L$, and $L_{2}=e/3\times L_{1}$, with $e$ the base of the natural logarithm.  
For comparison, the conventional estimate from \Eq{oscnom} is shown by the grey dashed line, evaluated at $E_{\g}=k\gamma^{2}$; varying $E_{\g}$ does not improve the agreement.  
Because the oscillations become extremely rapid, I omit the dashed curve for $m_{\g'}>1000\,k$ to reduce numerical cost.

Three qualitative differences emerge:

\begin{itemize}
  \item[{\bf a.}] \emph{Large‐mass damping.}  
    The conventional oscillatory behavior disappears at high $m_{\g'}$.
  \item[{\bf b.}] \emph{Reduced amplitude.}  
    The oscillation amplitude in our calculation is much smaller than the na\"{i}ve prediction.
  \item[{\bf c.}] \emph{Low‐mass enhancement.}  
    For small $m_{\g'}$ our result is enhanced relative to the standard formula.
\end{itemize}
These features originate from wave‐packet averaging and the full momentum/angle integration, effects absent in the one‐dimensional formula of \Eq{oscnom}. 
 
\paragraph{\bf Explanation of~{\bf a}.}  
The suppression at large mass arises from integrating over $k_{\g'}$ within the wave packet.  
For $L_{1}\neq L_{2}$ and very rapid oscillations, we can expand
\[
\frac{m_{\g'}^{2}}{4k_{\g'}}
  =\frac{m_{\g'}^{2}}{4\bar k_{\g'}}-\delta k_{\g'}\,\frac{m_{\g'}^{2}}{4\bar k_{\g'}^{2}},
\quad
k_{\g'}=\bar k_{\g'}+\delta k_{\g'}.
\]
Here $\bar k_{\g'}$ is the photon momentum in the wave packet and $\d k_{\g'}$ is the small deviation from it. Integrating first over $\delta k_{\g'}$ and then summing over $\bar k_{\g'}$ yields
\[
P_{\g\to\g}\;\longrightarrow\;
\chi^{4}\int\!\frac{d^{3}\bar k_{\g'}}{(2\pi)^{3}}
       \ab{\tl\F(\e,\g,\bar k_{\g'})}^{2}\times4,
\]
consistent with our numerical result. Here I assumed that $
       \ab{\tl\F(\e,\g, k_{\g'})}^{2}$ does not depend on $\d k_{\g'}$ much but the oscillation by varying $\d k_{\g'}$ is very fast. 
If $L_{1}=L_{2}$ so that the phases in the two sine factors coincide, the prefactor changes from $4$ to $6$, giving a slight enhancement.  Throughout this work I take $L_{1}\neq L_{2}$. This suppression by averaging the momentum mode is similar to the discussion given in \cite{Adler:2008gk}.

\begin{figure}[t!]
  \begin{center}  
\includegraphics[width=105mm]{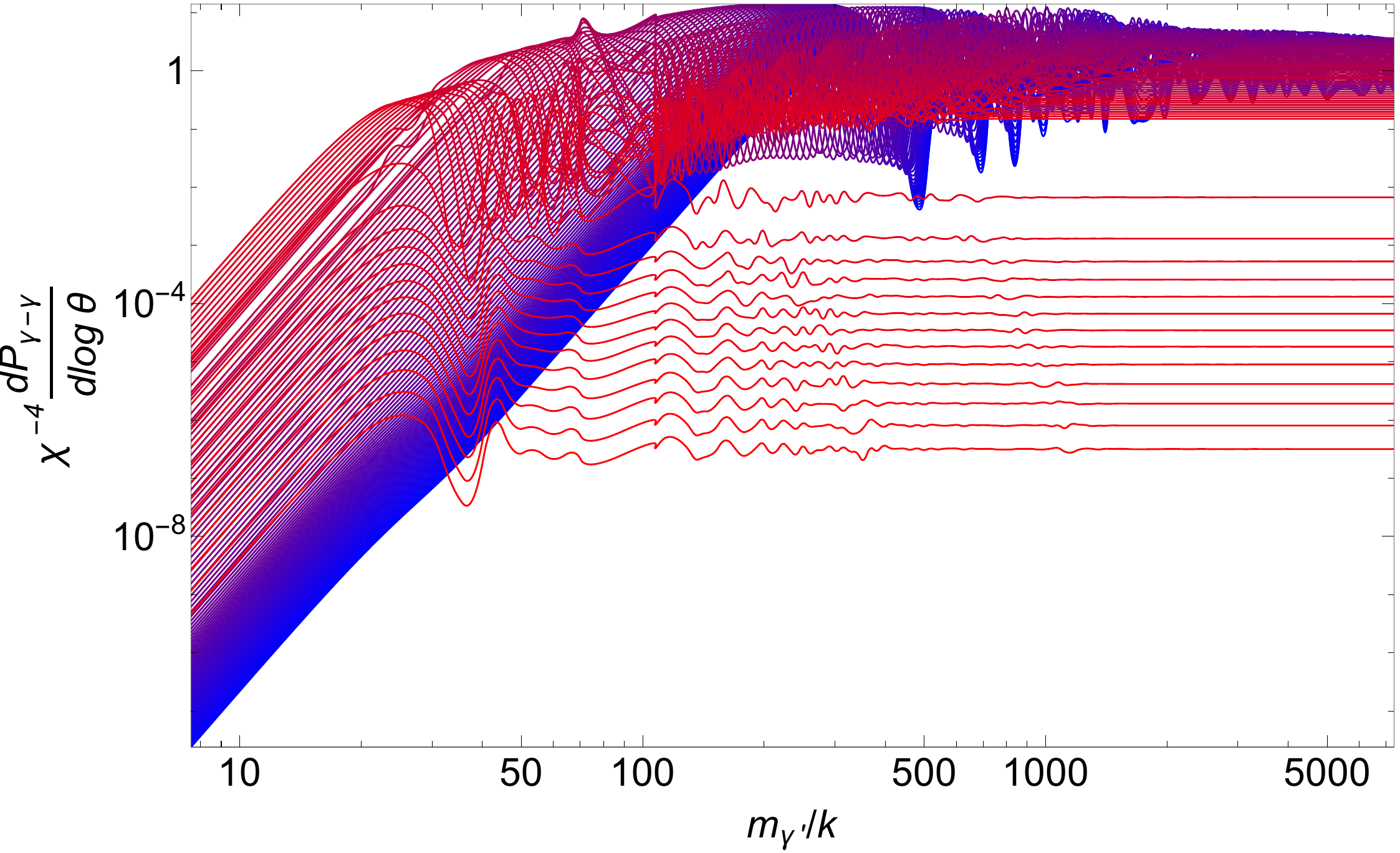}
  \end{center}
    \caption{$\partial_{\log_{10}[\theta]}P_{\g\to \g}$ by varying $m_{\g'}$ for the photon produced from undulator. 
 $\g=6\times 10^3, L=200/k$ are taken. $\log_{10} \theta$  is varied [-4.64282, -2.38282] with an interval of 1/50 from blue to red.  }
    \label{fig:osctheta}
    \end{figure}

\medskip
\noindent
{\bf Explanation of~{\bf b} and~{\bf c}.}  
Before integrating over the polar angle, the relevant quantity is the spectrum
$\partial_{\log_{10}\theta}P^{\rm eff, light}_{\g\to\g}$ shown in Fig.~\ref{fig:osctheta}, where
$\log_{10}\theta$ is varied from $-4.64$ to $-2.38$ in steps of $1/50$ (blue to red).

\begin{itemize}
\item[{\bf b.}]  At large $m_{\g'}$, each fixed‐$\theta$ mode oscillates rapidly, but summing over $\theta$ averages these oscillations out, reducing the overall amplitude.

\item[{\bf c.}]  In the small‐mass region the dominant contribution comes from
$\theta\simeq10/\gamma$, corresponding to photon energy $0.01\,k\gamma^{2}$ (see \Eq{kg}).  
Although this energy is well below the typical photon energy, the enhanced oscillation phase for this subset leads to less suppression in \Eq{oscnom}, boosting the transition probability by roughly two orders of magnitude at low $m_{\g'}$.
\end{itemize}

\subsection{A warm-up for (dark) photon particles interacting with wall and detector
$\displaystyle
 \frac{kL}{L_{1,2}}
 \ll \frac{m_{\g'}^{2}}{2k_{\g'}}
 \ll \frac{m_{\g,\rm p,pb}^{2}}{2k_{\g'}},\;\mu_{\rm pb}
$}
\lac{particleproduction}

At sufficiently large masses the oscillatory behavior disappears.
In the regime
$
\frac{kL}{L_{1,2}}
 \;\ll\;
 \frac{m_{\g'}^{2}}{2k_{\g'}},
% \ll\;
% \frac{m_{\g,\rm p,pb}^{2}}{2k_{\g'}},\;\mu_{\rm pb}
% ,
$
a particle picture applies: the undulator directly produces either the heavy eigenstate $\ket{2}$ or the massless eigenstate $\ket{1}$.  
Their production rates were computed in \Sec{dark photon} assuming the large $L_1$ limit. 

Thus the production stage is straightforward; the remaining task is to analyze how the mass eigenstates $\ket{2}$ (heavy) and $\ket{1}$ (massless) interact with the wall and subsequently couple to the detector. For clarity, I assume that $\mu_{\rm pb}, m^2_{\g,\rm p,pb}$ is so large that I neglect flavor oscillations inside the wall,  $ \frac{m_{\g'}^{2}}{2k_{\g'}}\ll \mu_{\rm pb}, \frac{m^2_{\g,\rm p,pb}}{2k_{\g'}} $.\footnote{If the reader is familiar with the dark photon and photon interacting with the wall neglecting microscopic structure, one may skip this subsection.}  This effect will be included in the following subsection.

Under this assumption the fractions $|\bra{\g'}1\rangle|^{2}\simeq\chi^{2}$ and $|\bra{\g'}2\rangle|^{2}\simeq1-\chi^{2}$ traverse the wall, while the complementary fractions  
$|\bra{\g}1\rangle|^{2}\simeq1-\chi^{2}$ and $|\bra{\g}2\rangle|^{2}\simeq\chi^{2}$ are absorbed.

Accordingly, for a photon produced by the undulator the probability that a dark photon flavor state $\g'$ emerges from the wall is  
\beq
P_{\g\to\g'}^{\rm eff,heavy}\;\simeq\;
\eta_{\rm eff}(m_{\g'})\,\ab{\bra{\g'}2\rangle}^{2}
+\ab{\bra{\g'}1\rangle}^{2}
\;\simeq\;
\bigl[\eta_{\rm eff}(m_{\g'})/\chi^{2}+1\bigr]\chi^{2},
\eeq
where $\eta_{\rm eff}$ is defined in \Eq{etaeff}.  
The first term comes from the heavy eigenstate produced in the undulator, projected onto $\g'$ at the wall; the second originates from the massless eigenstate.

On the detector side both massless and massive eigenstates propagate.  
Even when $m_{\g'}\gtrsim k\gamma$, so that the heavy mode is {not} directly produced in the undulator (see Fig.~\ref{fig:4}), it can still be generated inside the wall because $\mu_{\rm pb}\gg m_{\g'}^{2}/(2k_{\g'})$: energy non-conservation in the wall allows $\g'$ flavor to convert into the heavy eigenstate.  
This will be demonstrated explicitly with the kinetic equation in the next subsection.

The amplitudes reaching the detector are weighted by $|\bra{1}\g'\rangle|^{2}\simeq\chi^{2}$ and $|\bra{2}\g'\rangle|^{2}\simeq1-\chi^{2}$.  
Because the detector measures photon flavor, it responds with probabilities $|\bra{\g}1\rangle|^{2}\simeq1-\chi^{2}$ and $|\bra{\g}2\rangle|^{2}\simeq\chi^{2}$.  
Hence the normalized $\g'$ flavor state emerging from the wall is detected with probability  
\beq
\ab{\bra{\g'}1\rangle\bra{1}\g\rangle}^{2}
+
\ab{\bra{\g'}2\rangle\bra{2}\g\rangle}^{2}
\simeq
2\chi^{2}+\O(\chi^4).
\eeq
Interference terms are neglected because, in this mass range, the oscillation length is so short that the system can be treated in the incoherent particle picture as in the undulator side.
To sum up I estimate the effective transmission probability as
\beq
P_{\g\to \g}^{\rm eff,heavy}\simeq  \(\eta_{\rm eff}(m_{\g'})+\chi^2\)\times 2 \chi^2.
\eeq
%Here the factor of $2$ denotes the 

\subsection{Interference with the wall:
$\displaystyle
  \frac{kL}{L_{1,2}}
  \ll
  \frac{m_{\g'}^{2}}{2k_{\g'}}
$}

I now examine the effect of material interactions when the oscillation length becomes comparable to or shorter than the microscopic scales of the wall. 
For lead, the relevant mass scale obtained by equating $m_{\g'}^{2}/(2k_{\g'})$ to $\mu_{\rm pb}$ is
\beq
m_{\g'}\;\sim\;
30\EV
\sqrt{\frac{\mu}{2\times10^{-4}\;{\rm cm}^{-1}}\,
      \frac{k_{\g'}}{\KEV}}.
\eeq
This is much larger than the scale $k$ at which massive‐mode production in the undulator is already suppressed.  To keep the discussion general, I introduce a generic attenuation coefficient $\mu$ and plasma mass $m_{\g,p}$, postponing the specialization to lead until later. The effect discussed in this subsection includes the region discussed in the previous \Sec{particleproduction}.

\begin{figure}[t!]
  \begin{center}  
\includegraphics[width=100mm]{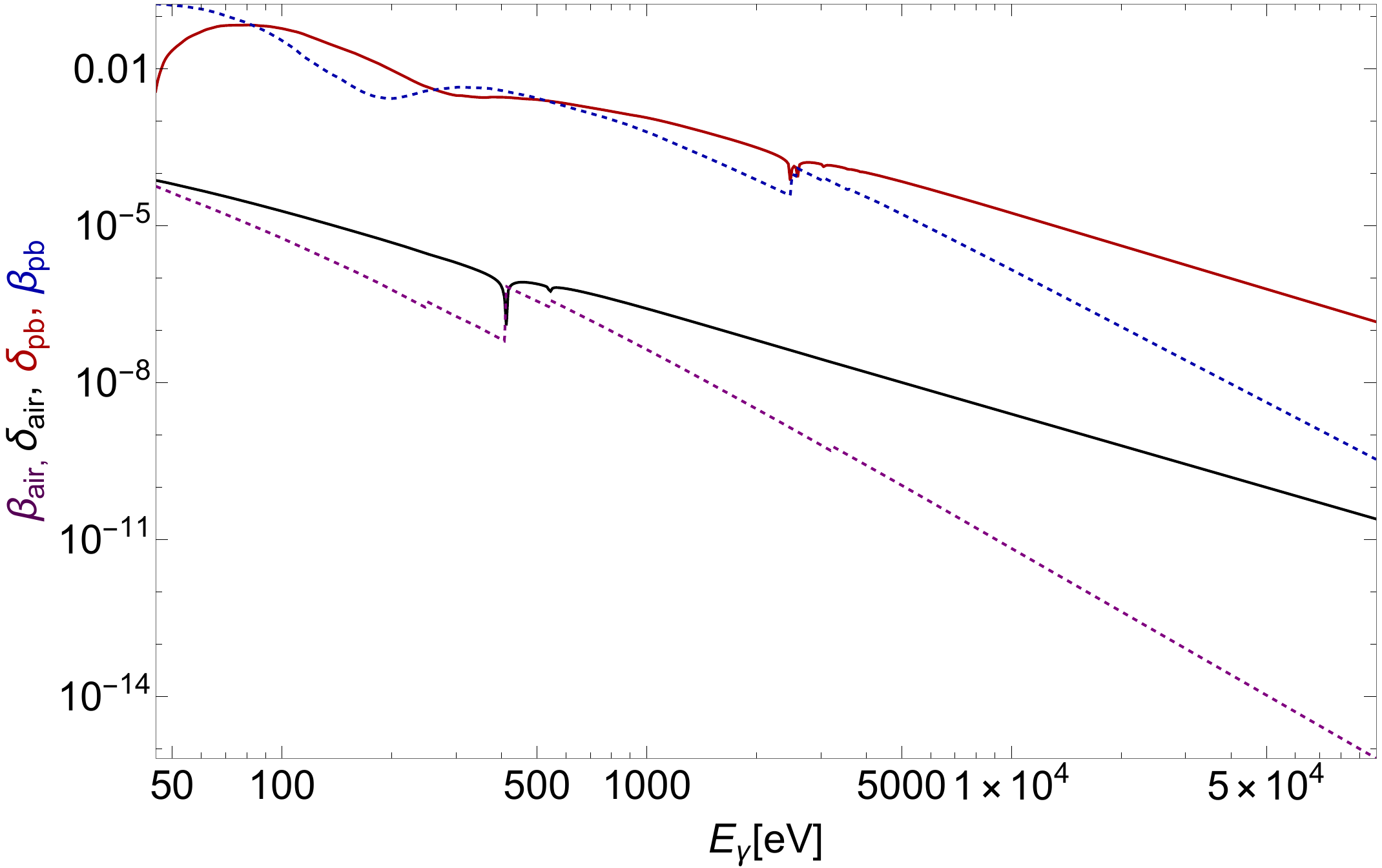}
  \end{center}
    \caption{
$\d$ and $\beta$ in the complex reflective indices of lead and air from top to bottom. 
I assume 11.34$\rm g/cm^3$ for the density of pb, and the air with the pressure of 1Pa and temperature of 295kelvin. They are estimated from \cite{henke1993x}.}
    \label{fig:muair}
    \end{figure}

The system is treated with the kinetic equation commonly used for neutrino oscillations in matter~\cite{Sigl:1993ctk},
\beq
\dot\rho=-i[H,\rho]-\frac12\{\Gamma,\rho\},
\eeq
where $\rho_{\a\b}$ is the density matrix in the mass basis for a mode with a momentum $k_{\g'}$.  The effective Hamiltonian and damping matrix are
\beq
H=
\begin{pmatrix}
  0 & 0\\
  0 & m_{\g'}^{2}/(2k_{\g'})
\end{pmatrix}
+
\begin{pmatrix}
  m_{\g,p}^{2}/(2k_{\g'}) & \chi\,m_{\g,p}^{2}/(2k_{\g'})\\
  \chi\,m_{\g,p}^{2}/(2k_{\g'}) & \chi^{2}m_{\g,p}^{2}/(2k_{\g'})
\end{pmatrix},
\qquad
\Gamma(k)=
\begin{pmatrix}
  \mu & \mu\chi\\
  \mu\chi & \mu\chi^{2}
\end{pmatrix}.
\eeq
Here, again $m_{\g,p}^{2}$ is the plasma mass squared of photons in the medium, and $\mu$ is the linear attenuation coefficient. 
Using the complex refractive index, $n\approx1-\delta(E_{\g})+i\beta(E_{\g})$, one finds
\beq
\laq{reflectiveindex}
\Pi_{T}
  =E_{\g}^{2}(1-n^{2})
  \approx
  2E_{\g}^{2}(\delta-i\beta)
  =m_{\g,p}^{2}-iE_{\g}\mu,
\eeq
where $\Pi_{T}$ is the transverse polarization tensor of the photon  in the medium.
  Tabulated values of $\delta$ and $\beta$ for various materials may be found in Refs.~\cite{henke1993x,chantler1995theoretical,chantler2000detailed}.%
\footnote{\url{https://henke.lbl.gov/optical_constants/}, 
\url{https://physics.nist.gov/PhysRefData/FFast/html/form.html}}
The specific cases for lead and for air, which will be used in the next section, are plotted in Fig.~\ref{fig:muair}. 

\begin{figure}[t]
  \centering
  \includegraphics[width=105mm]{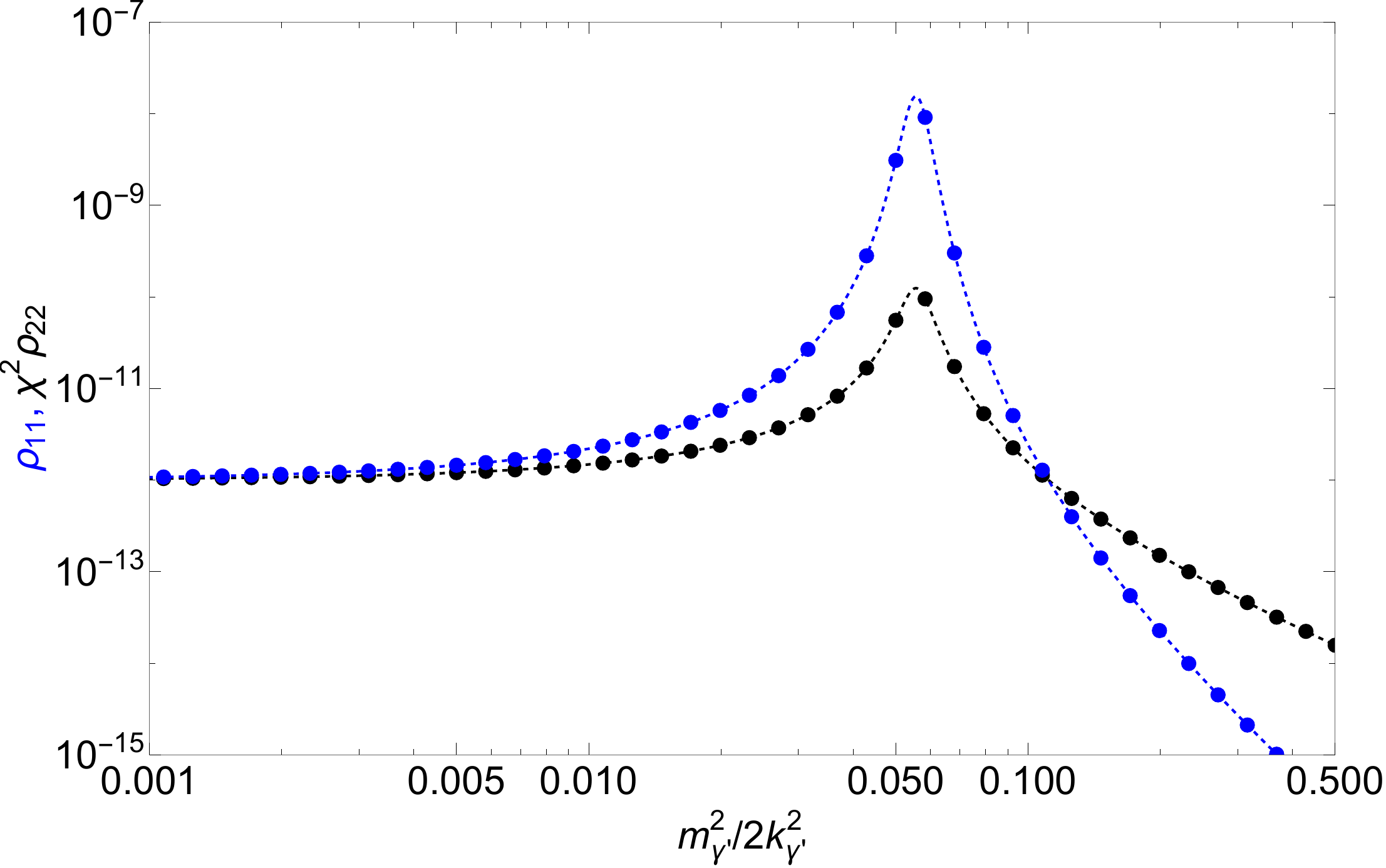}
  \caption{Numerical solution of the kinetic equation for the transmittance:
           $\rho_{11}$ (blue points) and $\chi^{2}\rho_{22}$ (black points) at
           $t=5000/k_{\gamma'}$.  Parameters:
           $\mu=k_{\g'}/100$, $m_{\g,p}=k_{\g'}/3$, $\chi=10^{-3}$, and
           $(\rho_{\alpha\beta})(t=0)=\delta_{\alpha1}\delta_{\beta1}$.  The
           dotted curve shows the analytic approximation in \Eq{ana}.}
  \label{fig:transmittance}
\end{figure}

For simplicity I take the initial condition to be the massless eigenstate,
\beq
(\rho_{\alpha\beta})(t=0)=\delta_{\alpha1}\,\delta_{\beta1},
\eeq
and defer other initial conditions to later discussion.
Fig.\,~\ref{fig:transmittance} displays the solution of the kinetic equation:
the transmission probabilities of the massless mode ($\rho_{11}$, blue points) and
the massive mode ($\rho_{22}$, black points) evaluated at
$t=5000/k_{\gamma'}$.
For the massive mode the factor $\chi^{2}$, the detection probability
$P_{\rm det}$, has been multiplied.
 I set the linear attenuation coefficient
$\mu=k_{\gamma'}/100$, the plasma mass $m_{\gamma,p}=k_{\g'}/3$, and the
mixing parameter $\chi=10^{-3}$.
One can see resonances at $m_{\g'}^2/(2k_{\g'})\sim m_{\g,p}^2/(2k_{\g'})$ for both modes and the modes get suppressed with $m_{\g'}^2/(2k)>\mu.$

The probabilities of the final states can be understood analytically via the complex Hamiltonian
\beq
\laq{QM}
H_{\rm eff}=
\begin{pmatrix}
  0 & 0\\
  0 & \dfrac{m_{\g'}^{2}}{2k_{\g'}}
\end{pmatrix}
+
\Bigl(\dfrac{m_{\g,p}^{2}}{2k_{\g'}} + i \dfrac{\mu}{2}\Bigr)
\begin{pmatrix}
  1 & \chi\\
  \chi & \chi^{2}
\end{pmatrix}.
\eeq
Its approximate eigenvectors are
\[
\phi_{\g',\rm eff}\;\approx\;
\bigl(
  \dfrac{\chi\bigl(-m_{\g,p}^{2}/k_{\g'}+i\mu\bigr)}
        {-i\mu+m_{\g,p}^{2}/k_{\g'}-m_{\g'}^{2}/k_{\g'}},
  1
\bigr),
\qquad
\phi_{\g,\rm eff}\;\approx\;
\bigl(
  1,
  \dfrac{\chi\bigl(m_{\g,p}^{2}/k_{\g'}-i\mu\bigr)}
        {-i\mu+m_{\g,p}^{2}/k_{\g'}-m_{\g'}^{2}/k_{\g'}}
\bigr),
\]
with eigenvalues $m_{\g'}^{2}/(2k_{\g'})$ and
$m_{\g,p}^{2}/(2k_{\g'})-i\mu/2$, respectively.
Thus $\phi_{\g',\rm eff}$ is the propagating mode, whereas
$\phi_{\g,\rm eff}$ is exponentially attenuated.  They coincide with the
flavor states $\ket{\g'}$ and $\ket{\g}$ only when
$\mu,\,m_{\g,p}^{2}/(2k_{\g'})\ll m_{\g'}^{2}/(2k_{\g'})$, i.e.\ in the regime analyzed in \Sec{particleproduction}.

Assuming the wall is thick enough to absorb the $\phi_{\g,\rm eff}$ mode, the
probabilities for a normalized initial state $\ket{1}$ to exit the wall as the
massless or massive eigenstate are
\begin{align}
\non
\ab{\langle1|\phi_{\g',\rm eff}\rangle\!\bra{\phi_{\g',\rm eff}}1\rangle}^{2}
  &\;\approx\;
  \ab{\frac{\chi\bigl(-m_{\g,p}^{2}/k_{\g'}+i\mu\bigr)}
          {-i\mu+m_{\g,p}^{2}/k_{\g'}-m_{\g'}^{2}/k_{\g'}}}^{4}
  \;=\;\chi^{4}\,
        \Bigl[
          \frac{m_{\g,p}^{4}+k_{\g'}^{2}\mu^{2}}
               {(m_{\g,p}^{2}-m_{\g'}^{2})^{2}+k_{\g'}^{2}\mu^{2}}
        \Bigr]^{2},\\[6pt]
\ab{\langle2|\phi_{\g',\rm eff}\rangle\!\bra{\phi_{\g',\rm eff}}1\rangle}^{2}
  &\;\approx\;
  \ab{\frac{\chi\bigl(-m_{\g,p}^{2}/k_{\g'}+i\mu\bigr)}
          {-i\mu+m_{\g,p}^{2}/k_{\g'}-m_{\g'}^{2}/k_{\g'}}}^{2}
  \;=\;\chi^{2}\,
        \frac{m_{\g,p}^{4}+k_{\g'}^{2}\mu^{2}}
             {(m_{\g,p}^{2}-m_{\g'}^{2})^{2}+k_{\g'}^{2}\mu^{2}},
\laq{ana}
\end{align}
respectively.
The analytic expressions in \Eq{ana} are plotted as the dotted curves in Fig.~\ref{fig:transmittance}; they agree remarkably well with the numerical solution.  
I also verify that  
\beq\ab{\langle1|\phi_{\g',\rm eff}\rangle\!\bra{\phi_{\g',\rm eff}}2\rangle}^{2}
 =\ab{\langle2|\phi_{\g',\rm eff}\rangle\!\bra{\phi_{\g',\rm eff}}1\rangle}^{2} \AND  
\ab{\langle2|\phi_{\g',\rm eff}\rangle\!\bra{\phi_{\g',\rm eff}}2\rangle}^{2}\!\simeq1,\eeq  
consistent with the kinetic equation result.  

Hence the effective transmission probability is
\beq \laq{hmass}
P_{\g\to \g}^{\rm eff,heavy} =\frac{1}{L\dot{n}_{\g}}\sum_{\e^{\pm}}\int \frac{d^3 k_{\g'}}{(2\pi)^3} \left\{ \ab{\langle{2,\e, \vec k_{\g'}|0_j\rangle}}^2   \chi^2     \(1+\frac{m^4_{\g,p}+  k^2_{\g'} \mu^2 }{  (m^2_{\g,p}- m^2_{\g'})^2+ k_{\g'}^2 \mu^2 }\)\right.
\ebq
+\left. \ab{\langle{1,\e, \vec k_{\g'}|0_j\rangle}}^2
\chi^4 \(\frac{m^4_{\g,p}+  k^2_{\g'} \mu^2 }{  (m^2_{\g,p}- m^2_{\g'})^2+ k_{\g'}^2 \mu^2 }+\Bigl[\frac{m^4_{\g,p}+  k^2_{\g'} \mu^2 }{  (m^2_{\g,p}- m^2_{\g'})^2+ k_{\g'}^2 \mu^2 }\Bigr]^2\)\right\}.
\eeq
The first term inside the braces represents the detection probability for the massive eigenstate originally produced in the undulator, while the second term corresponds to the massless eigenstate.  

\begin{figure}[t]
  \centering
  \includegraphics[width=105mm]{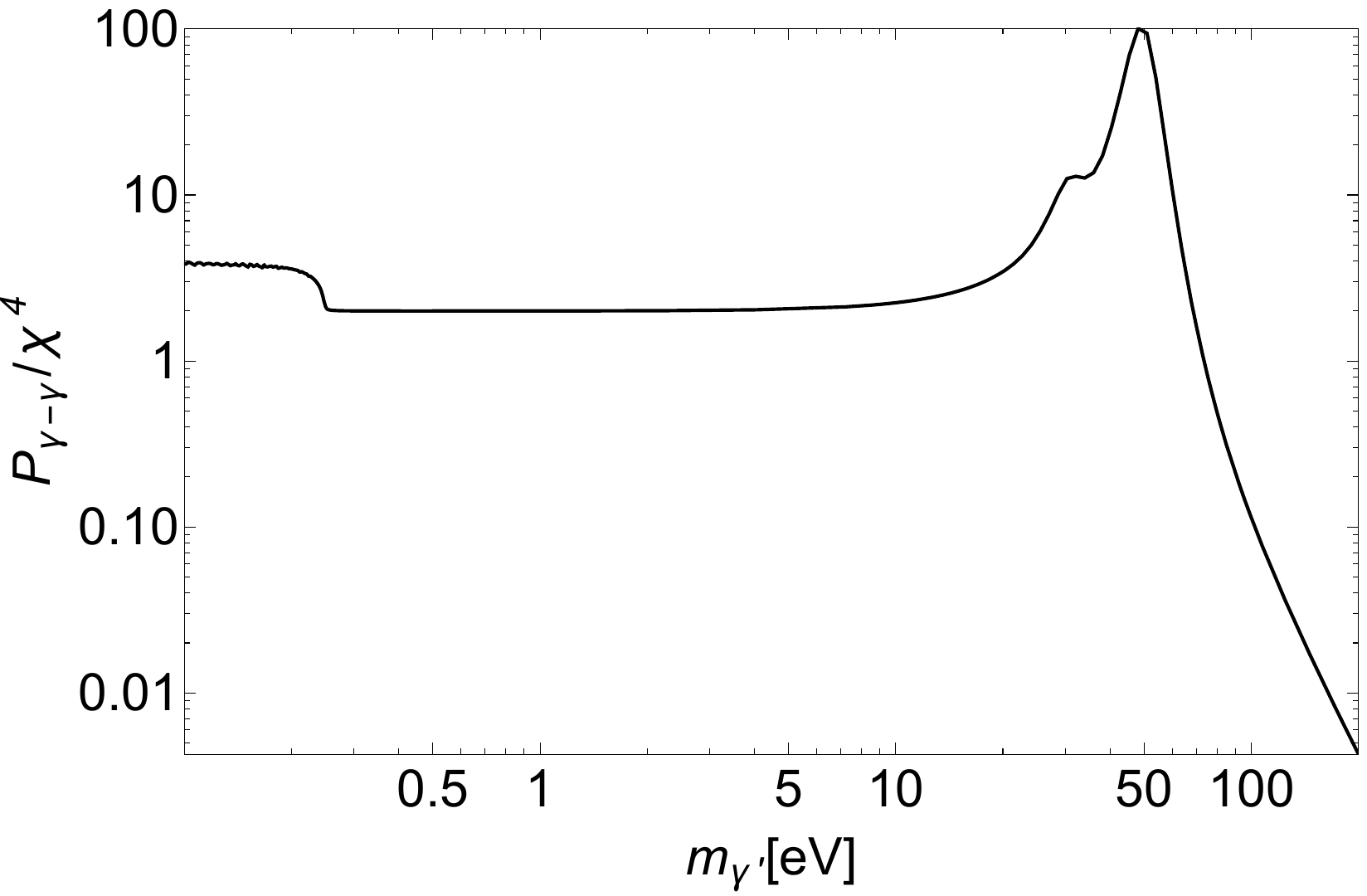}
  \caption{Same as Fig.~\ref{fig:osc}, but the heavy mass region is computed with the particle picture using QFT treatment, including the complex refractive index of the wall shown in Fig.~\ref{fig:muair}.}
  \label{fig:heavy}
\end{figure}

The numerically integrated result for $P_{\g-\g}^{\rm eff,heavy}/\chi^4$ is shown in Fig.\,\ref{fig:heavy} by assuming the setup given in Fig.\,\ref{fig:setup}. The suppression of the massive mode production reduce the photon detection rate by half around $m_{\g'}=0.2-0.3\EV$. The detection rate gets highly suppressed with $m_{\g'}\gg 50\EV$. 
The resonance effect with $m_{\g'}\sim 50\EV$ at which $m_{\g,p,\rm pb}\simeq m_{\g'}$ can be also seen to be efficient.

\subsection{Sensitivity limits: Undulators as dark photon factories}

I now estimate the prospective for the dark photon search.  The expected number of detected photons is  
\beq 
N_{\rm signal}=P^{\rm eff}_{\g\to\g}\,N_{\gamma}, 
\laq{signal}
\eeq 
where $P^{\rm eff}_{\g\to\g}$ is given by \Eqs{lmass} and \eq{hmass} in the light- and heavy-mass regimes, respectively.  These expressions coincide in the intermediate region $kL/L_{1,2}\ll m_{\g'}^{2}/k_{\g'}\ll k$, confirming the self-consistency of our treatment; accordingly, I switch from one formula to the other at a convenient mass within this overlap.

Assuming a photon-beam power of $15\,\mathrm{kW}$ and a typical photon energy $E_{\g}=k\gamma^{2}\simeq1.5~\mathrm{keV}$, values representative of modern synchrotron-radiation facilities, the primary photon rate is  
$
\dot N_{\gamma}\approx
\frac{15~\mathrm{kW}}{1.5~\mathrm{keV}}
\approx2\times10^{27}~\text{photons yr}^{-1}.
$
This is far larger than the photon rate actually delivered to experimental hutches, where mirrors, monochromators, and filters attenuate the beam; our calculation, however, concerns the \emph{primary} photons produced directly in the undulator.

Dark photons traverse the facility shielding and are detected outside the shield, which is assumed to be the lead in Fig.\ref{fig:setup}.\footnote{A more precise discussion for a particular synchrotron radiation facility, where we may have multiple ``walls'' which may be the mirror, filter, spectrograph etc, will be discussed in the future.  } With multi-year running, the enormous primary photon flux yields a projected sensitivity that surpasses existing laboratory limits in the relevant mass range by many orders of magnitude, although, like other LSW experiments, it still falls short of certain astrophysical and cosmological bounds (see, e.g., Ref.~\cite{Caputo:2021eaa}).

\section{LSW in the air}
\lac{air}

Up to this point I have assumed a perfect vacuum everywhere.  A simpler and experimentally more economical configuration places the detector in ambient air while keeping the undulator section under ultra-high vacuum.  The latter is essential to preserve electron‐beam quality and to prevent keV photons from being attenuated before they are used for experiments; treating the undulator region as vacuum, which is the case for many synchrotron radiation facilities, is therefore well justified.  All other experimental parameters remain unchanged, except that the detector, located outside the photon shield, now sits in air.

Air has a sizeable linear attenuation coefficient,
$\mu_{\rm air}\sim10\;\mathrm{cm^{-1}}$ at keV energies, so most photons are absorbed before they can traverse even modest path lengths.  In practice the air volume itself functions as an additional “wall.’’  
The real ($\delta$) and imaginary ($\beta$) parts of the complex refractive index are plotted in Fig.~\ref{fig:muair} for standard atmospheric pressure and $T=293~\mathrm{K}$.  
Under these conditions the photon-like eigenstate produced at the shield cannot reach the detector; only the dark photon-like eigenstate can propagate through the air.

\subsection{Quantum Zeno effect due to the medium}

\begin{figure}[t]
  \centering
  \includegraphics[width=105mm]{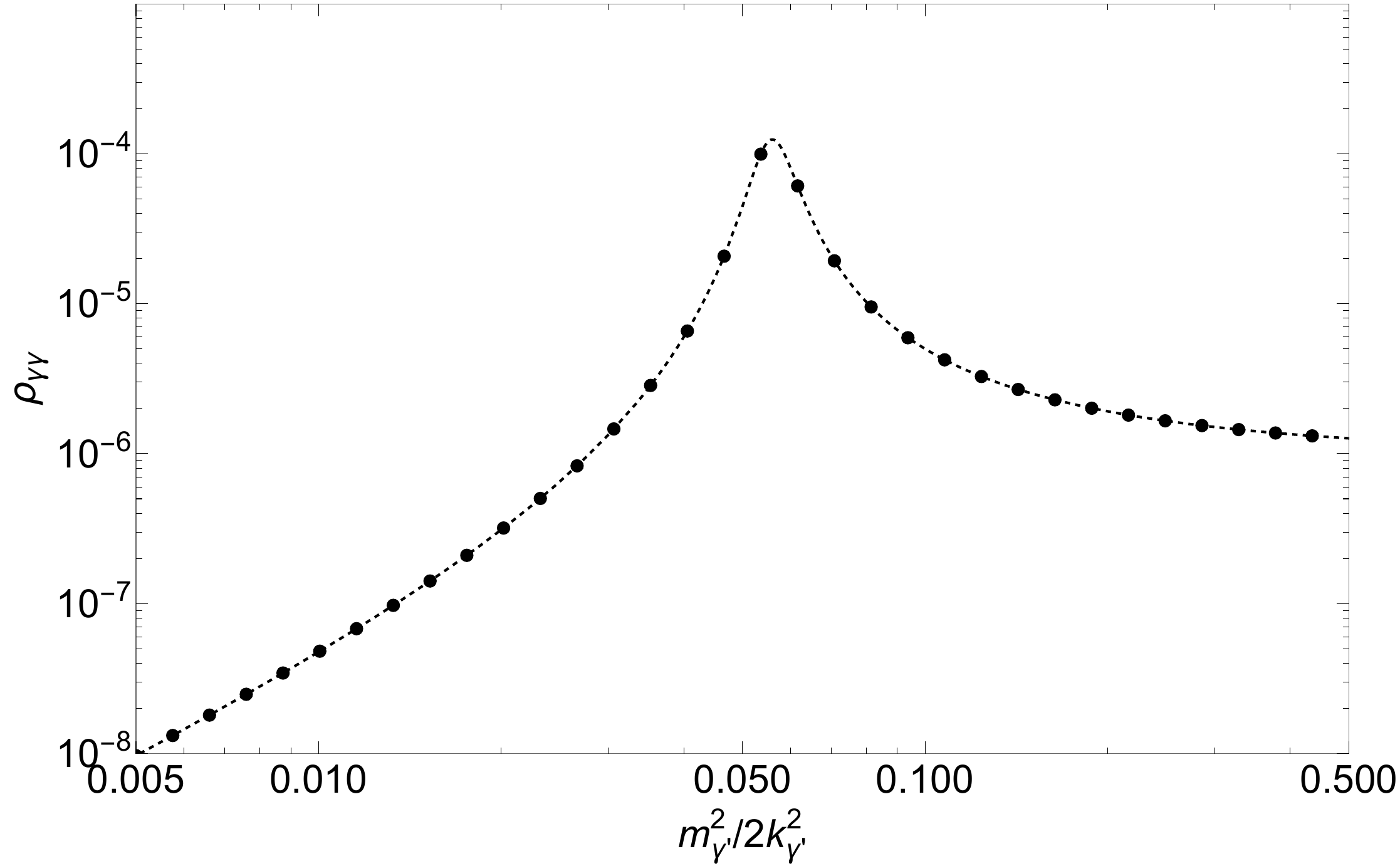}
  \caption{Solution of the kinetic equation for the photon‐flavor transmittance
           $\rho_{\g\g}$ at $t=5000/k_{\gamma'}$ (black dots).
           Parameters:
           $\mu=k_{\g'}/100$ and $\chi=10^{-3}$;
           initial condition $(\rho_{ij})(0)=\delta_{i\g'}\delta_{j\g'}$.
           The dotted curve is the analytic estimate in \Eq{ana2}.}
  \label{fig:quantumzeno}
\end{figure}

To assess the impact of air, I again solve the kinetic equation. Here the initial state is the dark photon flavor that has just emerged from the wall:
\beq
(\rho_{ij})(t=0)=\delta_{i\g'}\delta_{j\g'}.
\eeq
I then monitor $\rho_{\g\g}$ at sufficiently late times to quantify the coupling to the detector.  
Fig.\,~\ref{fig:quantumzeno} shows $\rho_{\g\g}(t=5000/k_{\gamma'})$ (black dots).  
All parameters match those in Fig.~\ref{fig:transmittance}.

In fact it is known that the propagating state in the medium couples to the charged particle with an effective mixing
(e.g.\ \cite{An:2013yua,An:2014twa}):\footnote{The same expression can be derived by diagonalizing the complex Hamiltonian in \Eq{QM}.}
\beq
\chi_{\rm eff}^{2}\;\equiv\;\ab{\bra{\g}\f_{\g',\rm air}\rangle}^{2}
\simeq
\chi^{2}\,
\frac{m_{\g'}^{4}}
     {(m_{\g'}^{2}-m_{\g,p}^{2})^{2}+k_{\g'}^{2}\mu^{2}},
\laq{ana2}
\eeq
where $\ket{\f_{\g',\rm air}}$ denotes the dark photon-like propagating mode in
air (valid when $\chi^{2}\mu L_{2}\ll1$).  The suppression, when $m_{\g'}$ is small, can be understood by the quantum Zeno effect (see e.g. \cite{Sakurai:2022cki}). 
Because frequent “measurements’’ by the medium project out the photon component, $\ket{\f_{\g',\rm air}}$ is almost a pure dark photon flavor rather than a mass eigenstate.  
Again, there is a resonant enhancement when $m_{\g'}\sim m_{\g,p}$. 
The dashed line in Fig.~\ref{fig:quantumzeno} shows the analytic formula \Eq{ana2}, which agrees excellently with the numerical solution of the kinetic equation.

There is a useful relation between the propagating modes in different media,
\beq 
\ab{\bra{\f_{\g',\rm eff}}\f_{\g',\rm air}\rangle}^2 \approx 1,
\eeq
since $\ab{\bra{\f_{\g',\rm eff}}{\g'}\rangle}^2\approx \ab{\bra{\f_{\g',\rm air}}{\g'}\rangle}^2\approx 1$. This implies that for a sequence of media that attenuate the photon flavor state, the intermediate layers do not affect the final transition probability. In each medium, the propagating mode is dominantly the dark photon-like state.

   \begin{figure}[t!]
  \begin{center}  
\includegraphics[width=70mm]{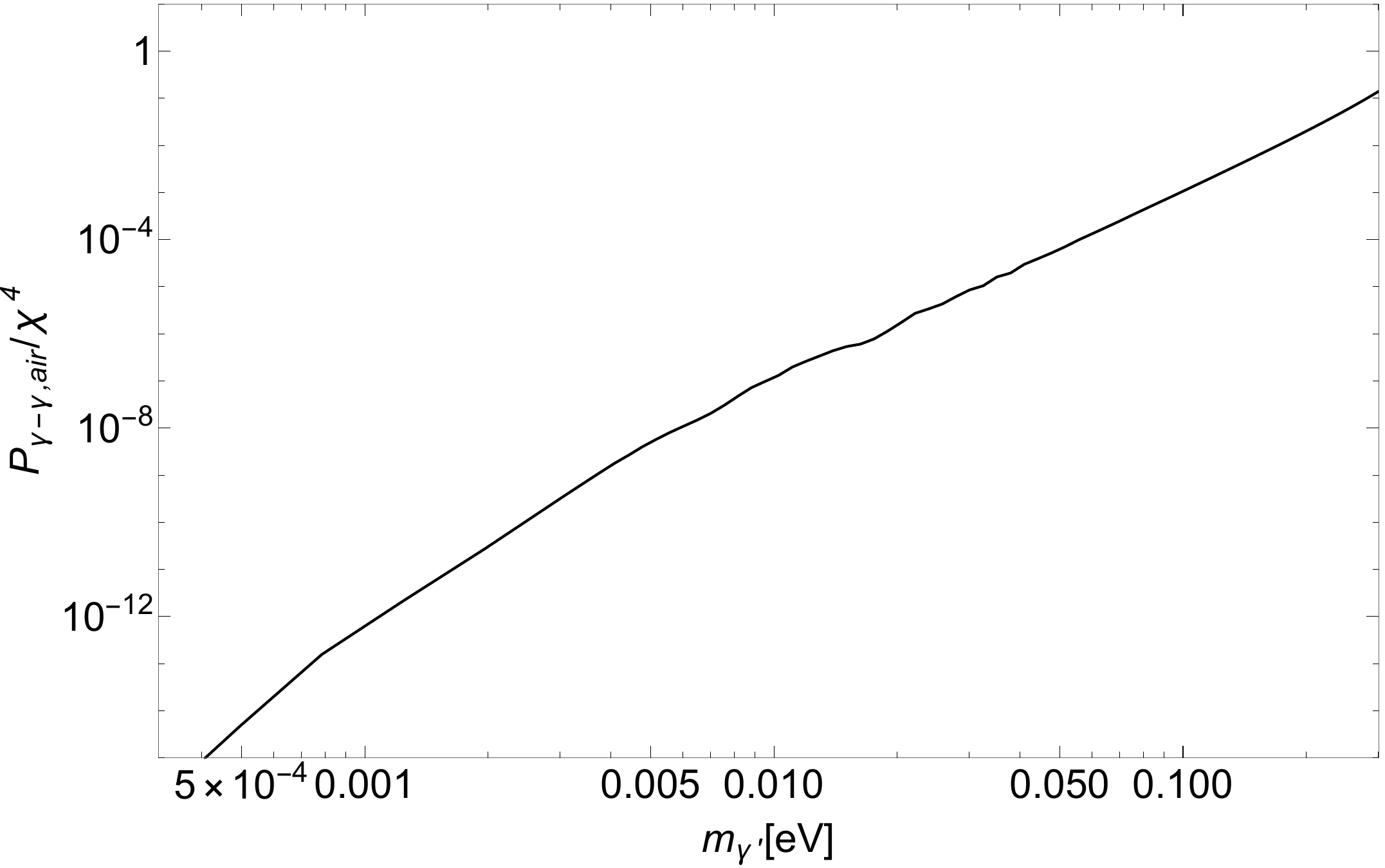}\hspace{3mm}
\includegraphics[width=70mm]{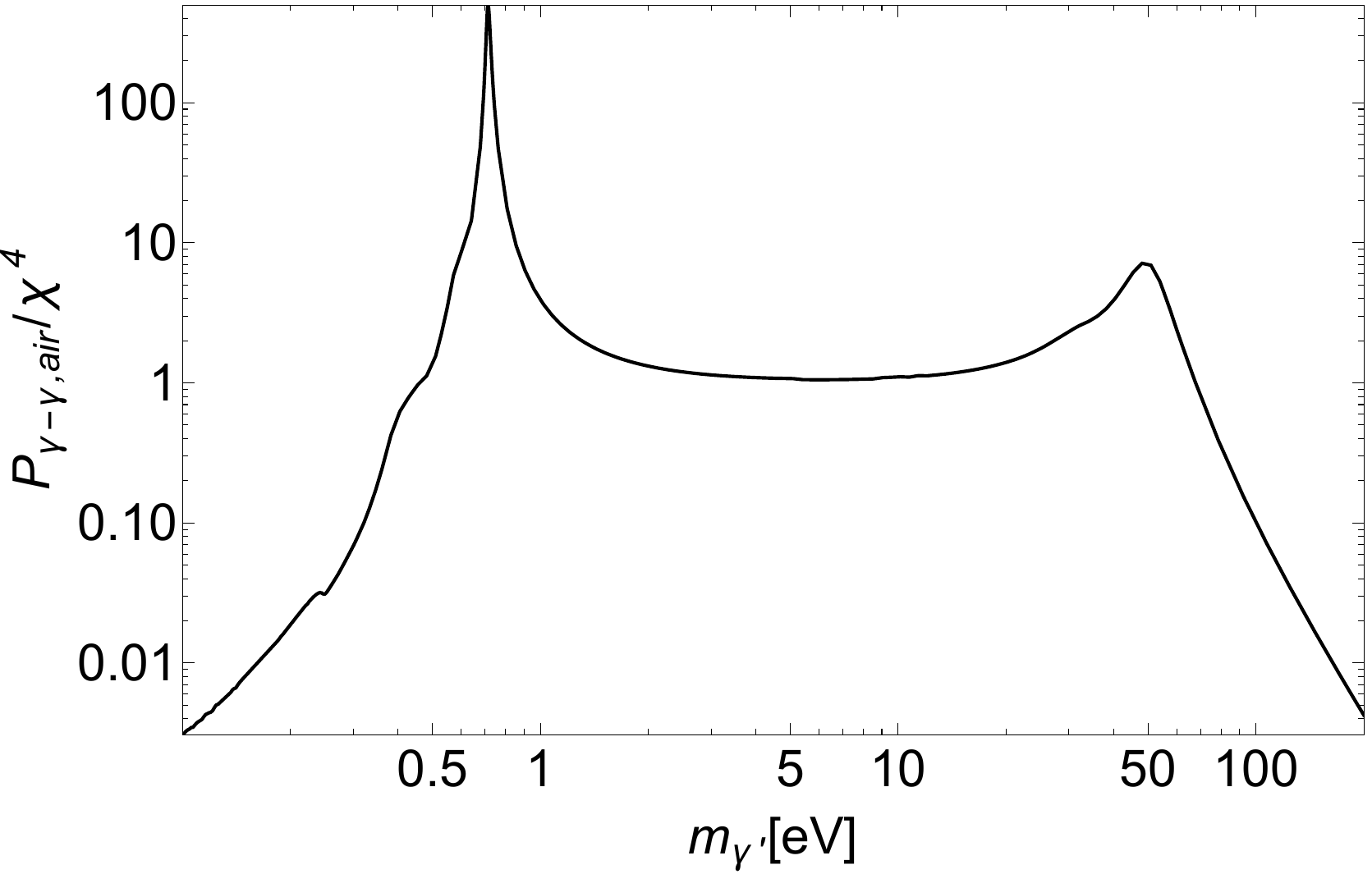}
  \end{center}
    \caption{Same as Figs.\,\ref{fig:osc} (left panel) and \ref{fig:heavy} (right panel) but the detector side is filled with air. The effects of air attenuation and quantum zeno are included with the reflective index in Fig.\ref{fig:muair}.}
    \label{fig:oscair}
    \end{figure}

\subsection{Future sensitivity with the detector side in air}

Following the discussion in \Sec{qosc}, I now evaluate the transmittance probability in the light mass regime with the effect of air.  The result can be written
\beq
P_{\g\to\g,\,\rm air}^{\rm eff,light}
\;\approx\;
\frac{1}{L\dot n_{\g}}\,
\chi^{4}
\sum_{\e^{\pm}}
\int\!\frac{d^{3}\vec k_{\g'}}{(2\pi)^{3}}\;
\ab{\tl\F(\e,\g,\vec k_{\g'})}^{2}\,
4\sin^{2}\!\Bigl(
  \frac{m_{\g'}^{2}}{4k_{\g'}}\,L_{1}
\Bigr)\;
\frac{m_{\g'}^{4}}
     {(m_{\g'}^{2}-m_{\g,p,\rm air}^{2})^{2}
       +\mu_{\rm air}^{2}k_{\g'}^{2}},
\laq{lmassair}
\eeq
where the air parameters $m_{\g,p,\rm air}$ and $\mu_{\rm air}$ are defined in Fig.\;\ref{fig:muair}.  
Relative to \Eq{lmass} there are two crucial differences:

\begin{enumerate}
\item Because ordinary photons are heavily attenuated in air, no flavor oscillation occurs after the wall.  
      Only the dark photon like mode propagates to the detector.
\item The effective mixing angle between the propagating mode and the detector 
      is introduced following \Eq{ana2}.  
\end{enumerate}

Integrating over momentum yields the curve shown in Fig.\;\ref{fig:oscair} (left panel) for small $m_{\g'}$.  The right panel displays the corresponding heavy-mass result, obtained with the analogue of \Eq{hmass} that incorporates both air attenuation and the effective mixing to the detector.

For the heavy mass regime, I employ
\begin{align}
 \laq{hmassair}
P_{\g\to \g,\rm air}^{\rm eff,heavy} &=\frac{1}{L\dot{n}_{\g}}\int \frac{d^3 k_{\g'}}{(2\pi)^3} \sum_{\e^{\pm}}\left\{\ab{\langle{2,\e, \vec k_{\g'}|0_j\rangle}}^2 \chi^2 
+
\non
\ab{\langle{1,\e, \vec k_{\g'}|0_j\rangle}}^2
 \chi^4 \frac{m^4_{\g,p,\rm pb}+  k^2_{\g'} \mu_{\rm pb}^2 }{  (m^2_{\g,p,\rm pb}- m^2_{\g'})^2+ k_{\g'}^2 \mu^2_{\rm pb}}
\right\}\\ & ~~~~~~~~~~~~~~~~~~~~~~~~~~~\times  \frac{ m_{\g'}^4}{(m_{\g'}^2-m_{\g,p,\rm air}^2)^2+\mu_{\rm air}^2 k_{\g'}^2}.
\end{align}
Relative to \Eq{hmass} the photon-like propagation term has been removed and the
air‐induced effective mixing \Eq{ana2} to the detector is incorporated.

\begin{figure}[t!]
  \begin{center}  
\includegraphics[width=145mm]{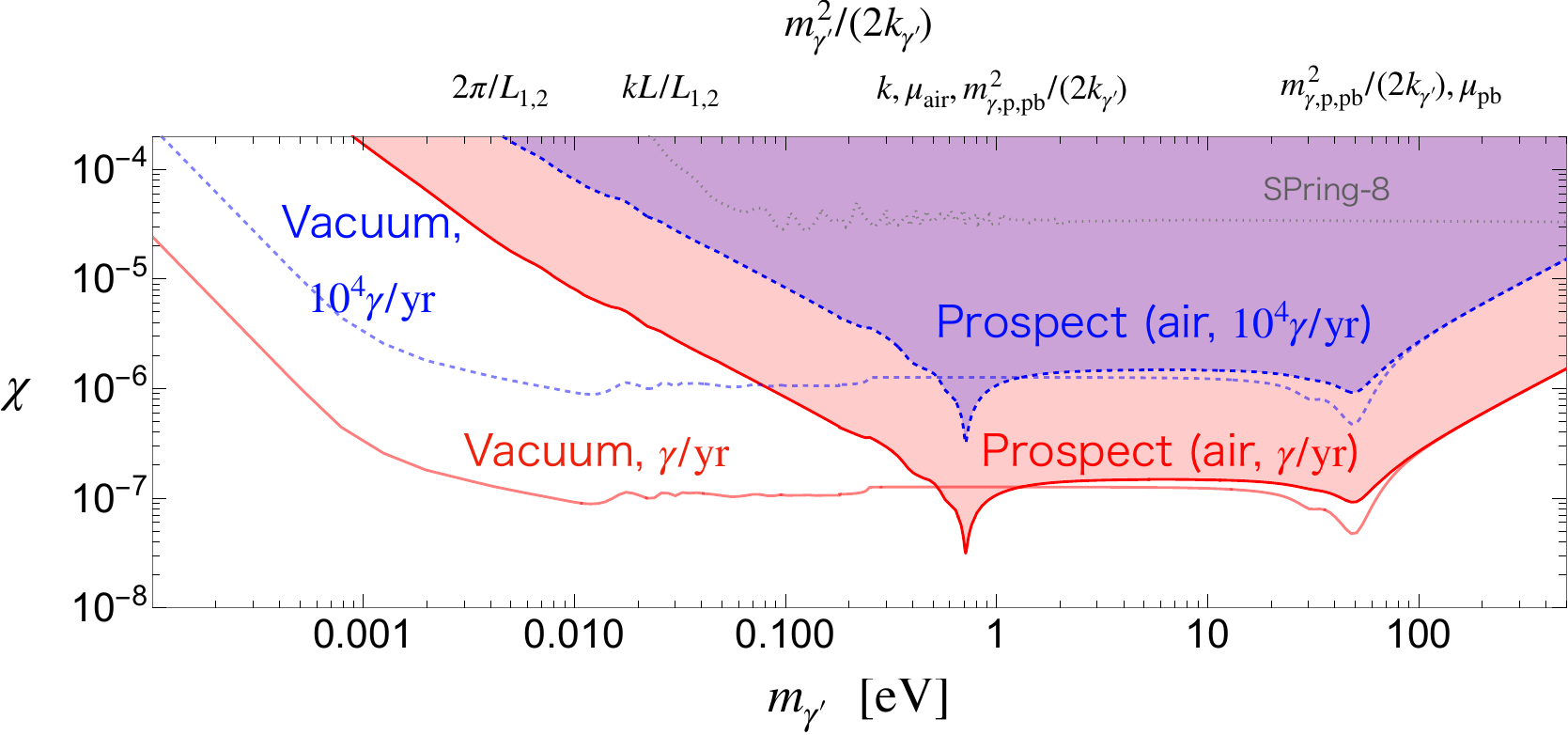}
  \end{center}
    \caption{Same setup as Fig.~\ref{fig:limit} but the sensitivity limit is for the detector side filled with air. For comparison the corresponding sensitivity limit of  Fig.~\ref{fig:limit} is also shown in light red solid and light blue lines. 
    }
    \label{fig:limitair}
    \end{figure}

Combining the light- and heavy-mass regions and using \Eq{signal},
I obtain the sensitivity reach plotted in Fig.~\ref{fig:limitair}.  
Except for filling the detector side with air, all parameters match those in
Fig.~\ref{fig:limit}.  
Compared with the vacuum case, the sensitivity degrades markedly for
$m_{\g'}\ll1\EV$ owing to the quantum Zeno suppression, whereas for
$m_{\g'}\gtrsim1\EV$ the reach is only mildly affected and still improves upon
existing laboratory bounds.  
Intriguingly, a pronounced enhancement appears near $m_{\g'}\sim1\EV$.\footnote{This is
more relevant with larger $k \g^2$, where the imaginary part of the refractive
index exceeds the real part (see Fig.~\ref{fig:muair}). Thus for an undulator with more energetic photon the enhancement can be more significant.}

Thus, even without evacuating the detector volume, a compelling dark photon
search is possible and for certain masses the air-filled configuration
outperforms the vacuum case.  In practice, simply installing a photon detector
outside the synchrotron shield along the beamline could tighten LSW limits
significantly.

Finally, consider enclosing the detector in a finite vacuum box within the air
volume.  Inside the box photon-dark photon oscillations resume, and the mass
eigenstates couple to the detector with unsuppressed amplitudes $1$ and $\chi$.
If the box is long enough for oscillations to develop, the low-mass sensitivity
improves, interpolating between the vacuum and air curves in
Fig.~\ref{fig:limitair}.  A detailed optimization of box size and placement is
left for future work with a concrete facility.

\section{Conclusions and discussion}
\lac{conclusion}
In this work I have revisited light-shining-through-a-wall (LSW) searches for
dark photons in which the light source is an undulator.  In contrast to the
conventional treatment based solely on the na\"{i}ve photon-dark photon
oscillation formula, I have incorporated three key quantum-mechanical effects:

\begin{enumerate}
\item finite wave-packet size,
\item kinematical suppression originating from the microscopic structure of the source, and
\item medium-dependent mixing in both the wall and the surrounding atmosphere.
\end{enumerate}

Each of these effects can appreciably modify the sensitivity of an LSW
experiment.  In particular, with the realistic complex refractive index of the wall (and of
air) the medium effect introduces resonance-like features that either enhance or suppress the
signal, depending on $m_{\g'}$.  Sensitivity is therefore \emph{not} universal
but is highly dependent on the experimental environment and the propagation
conditions of photons and dark photons. This study demonstrates that a fully quantum-field-theoretic analysis of the
\emph{complete} system, source, wall, and medium, is indispensable for reliable
predictions and for designing realistic, high-sensitivity searches for dark
sector particles.

Accounting for all of these effects, I have shown that a \emph{parasitic}
search is feasible: a photon detector placed outside the shielding, along the
beamline of an existing synchrotron facility, can exploit the highly
collimated undulator light without disrupting normal operation.  Such a setup
is economical and practical, and it probes hitherto-unexplored regions of dark
photon parameter space in laboratory.\\

Although I concentrated on dark photons, the same quantum effects—medium-induced
mixing, decoherence due to large wave packets, and kinematical suppression in
particle production—have broad applicability, for example to neutrino
oscillations in matter (with BSM interactions) and to ALP-photon
conversion.

Finally, I comment the caveat {\bf 4.} in the Introduction. So far, I have assumed an idealized photon detector.  In practice, detector
response depends on the specific detection mechanism and can modify
sensitivity, especially for the heavy eigenstate, where the na\"{i}ve $\chi^{2}$
scaling may be overly conservative.   A realistic detector has finite thickness: the photon‐like component may be absorbed in the first few layers, whereas the dark photon-like component, being weakly mixed, can traverse deeper layers and still reconvert to photons. This effect as well as the resonant enhancement of the mixing due to the medium can increase the overall detection probability.  A detailed detector‐level analysis will be presented in an upcoming article where I will show that even a simple Geiger counter operated under standard radiation‐safety conditions at a synchrotron facility can set some of the strongest ground‐based limits on dark photons in certain mass range.

\section*{Acknowledgments}
W.Y. thank the useful discussions with Junya Yoshida, T. Moroi, and  Toshio Namba. 
This work is supported by JSPS KAKENHI Grant Numbers  22K14029 (W.Y.), 22H01215 (W.Y.) and Tokyo Metropolitan University Grant for Young Researchers (Selective Research Fund).

\clearpage

\appendix

\section{Behavior at $\theta \approx 0$ and the resonance in the massive mode production from undulator}
\begin{figure}[t!]
  \begin{center}  
\includegraphics[width=140mm]{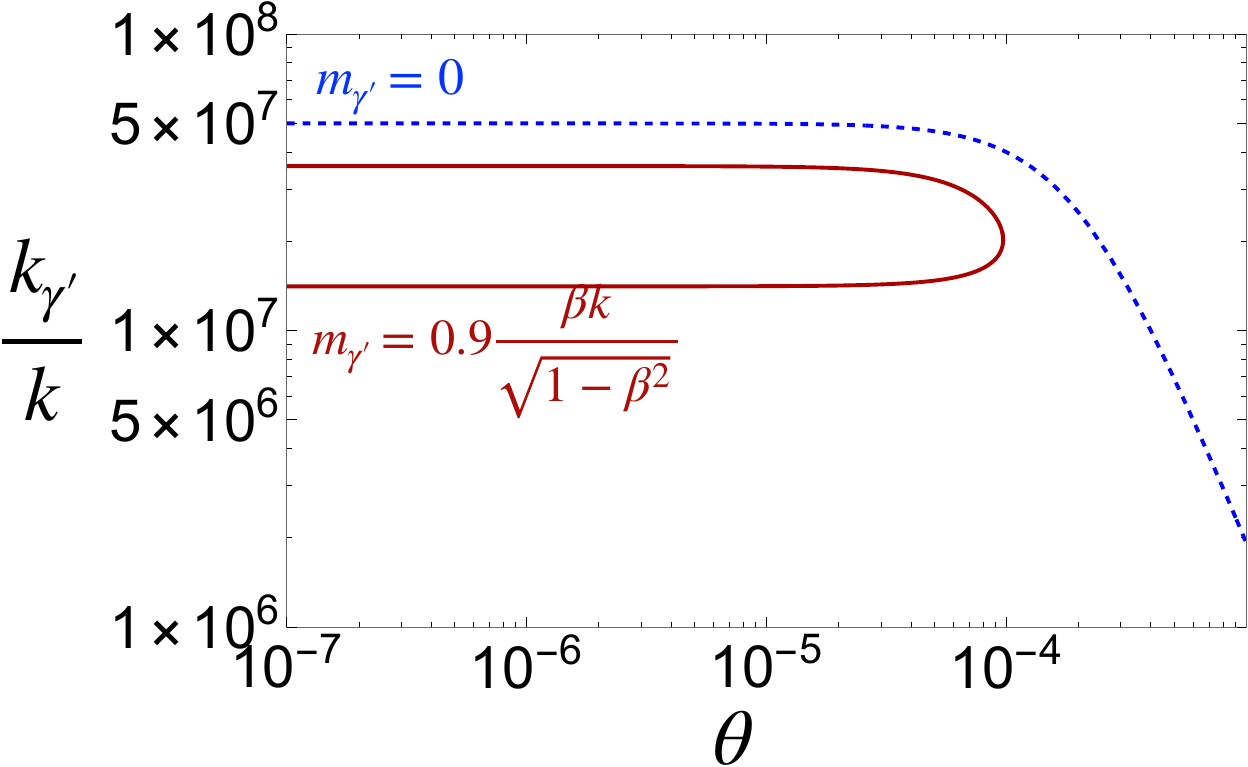}
  \end{center}
    \caption{Dominant modes of $k_{\g'}/k$ by varying $\theta$. The blue dashed line corresponds to $m_{\g'}=0$ (the photon case), and the red solid line to $m_{\g'}=0.9\frac{\beta k}{\sqrt{1-\beta ^2}},$ with Lorentz factor $\gamma=5000$.}
    \label{fig:Multi}
    \end{figure}

When the delta-like function's argument of \Eq{delta} is zero, I obtain
 \beq
k_{\g'}= \bar k_{\g'}\equiv \frac{ \pm \sqrt{4 \beta ^2 k^2+2\beta ^2 m_{\gamma'}^2 \cos2 \theta +2\beta ^2 m_{\gamma'}^2-4 m_{\gamma'}^2}+2 \beta ^2 k \cos \theta }{2 \left(1-\beta ^2 \cos ^2\theta \right)},
 \eeq
 which gives the dominant mode for the production. 
 Interestingly, this is a multi-value function for $m_{\g'}\neq 0$. An example with $\gamma=5000$ is shown in Fig.\ref{fig:Multi}. 
 The multi values agree with each other when 
 \beq\laq{f'zero}
 m_{\gamma' }\simeq \frac{\beta  k}{\sqrt{1-\beta^2 \cos^2\theta}}.
 \eeq

\begin{figure}[t!]
  \begin{center}  
\includegraphics[width=140mm]{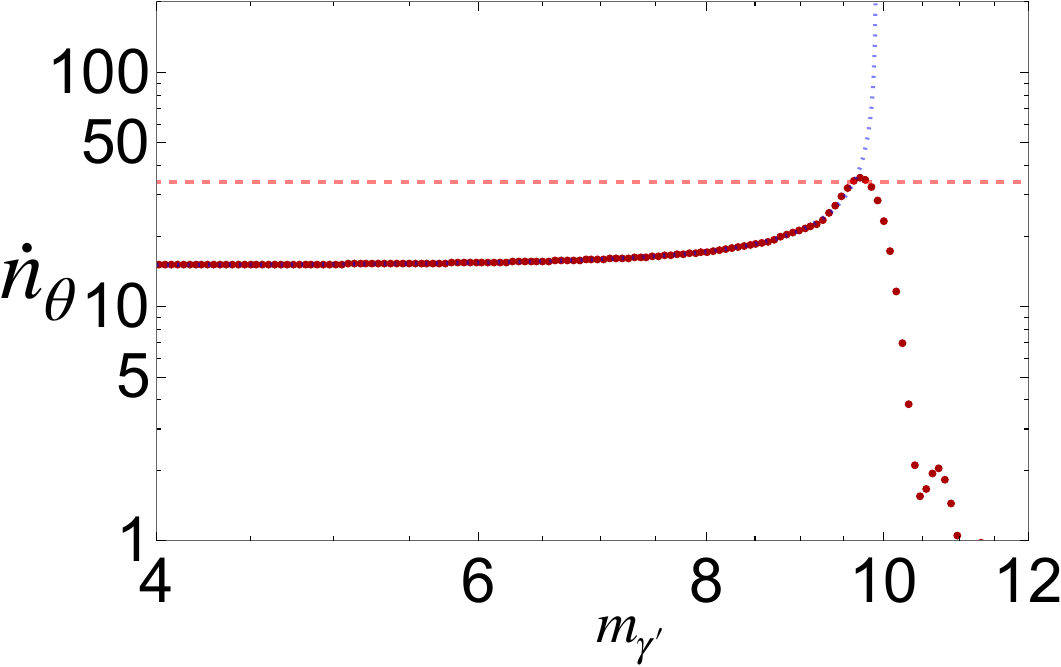}
  \end{center}
    \caption{$\dot n_\theta$ by varying $m_{\g'}$. $k=1,\theta=0.01,L=100, \gamma=10$}
    \label{fig:2}
    \end{figure}

$
\dot n_\theta
$ 
can be numerically evaluated as the red data points in Fig.\ref{fig:2} by using $\delta_L$  in \Eq{delta}. Here I take $k=1,\theta=0.01,L=100$. 
 Taking $\theta=0$ does not change the figure much. 
 One can see that when $m_{\g'}$ increases to the value $\sim \frac{ k}{\sqrt{1-\beta ^2 }},$ there is an enhancement and then the flux gets suppressed. 
 The blue dotted line and horizontal red dashed line are the analytical estimations to explain the behavior that are discussed below. 

\paragraph{\textcolor{blue}{Blue dotted line in Fig.\ref{fig:2}}}
At the leading order, and $\theta \simeq 0$ I get 
\begin{align}\laq{J}
 \dot n_{\theta} \simeq&\left.\frac{ 1}{|f'|}\frac{e^2K^2 k_{\gamma'}^4 }{8 \pi  \gamma ^2 k^2 w_{\g'}}\right|_{k_{\g'}=\bar k_{\g'}}.
%\\ 
%w_{\g'} &=\frac{\b k}{1-\cos\theta\beta} 
%\laq{wg}
\end{align}
Here, I used 
$\int_{-\infty}^{\infty}{d x\ab{\pi\delta_L(f(x))}^2}=2\pi L /|f'|$ which is the usual formula for the delta function, with 
\beq
f(k_{\g'})=\partial_{k_{\g'}}(\D k_{\gamma'} - k \beta)= \frac{k_{\g'}}{w_{\g'}}-\b \cos\theta.
\eeq 
This formula with not too large $m_{\g'}$, agrees very well with the numerical result as can be seen in Fig.\ref{fig:2}. 
In particular, when $m_{\g'}=0$ the formula agrees with the one for the usual synchrotron radiation. 

The difference from the usual photon case is that $f'$ can be very small when $m_{\g'}$ is non-vanishing. Due to the factor of $1/|f'|$ one gets enhancement of flux. 
This enhancement is towards the critical $m_{\g'}$ satisfying $f'=0, f=0$ which gives
\Eq{f'zero} which is the value that multi-value function of $k_{\g'}$ becomes the single one. With this choice the analytic formula diverges.

\paragraph{\textcolor{red}{Horizontal red dashed line in Fig.\ref{fig:2}}} 

The divergence of the blue dotted line is an artifact of our analysis because when $f'$ is small we need to take into account of the next leading term in the argument of the `delta function'. 

The expansion around the zero value we have \beq 
f(x)\approx f'(x_0) \Delta x + f''(x_0) \frac{\Delta x^2}{2} \cdots ,\eeq 
where $f(x_0)=0, \D x = x-x_0$.
%we need to neglect the second order. 

%Then we get $\D x\ll  f'(x_0)/f''(x_0)$, while we also have the validty condition of using the delta-like function as delta function, $1/L\lesssim |f'(x_0) \Delta x + f''(x_0) \Delta x^2/2|.$
When $f'\to 0$, $f''$ term dominates and I have  
\beq
\int dx |\delta_L(f(x))|^2\sim L\sqrt{L /|f''|}\sim L \sqrt{L k^3_{\g'}/m^2_{\g'}}\sim L \gamma^2 \sqrt{L k} .
\eeq
Thus the analytic formula \Eq{J} should be cutoff at this value.

\beq
 \dot n_{\theta} |_{f'=0}\sim \left.\sqrt{\frac{L}{|f''|}} \frac{1}{2\pi} \times \frac{e^2K^2 k_{\gamma'}^4 }{8 \pi  \gamma ^2 k^2 w_{\g'}}\right|_{k_{\g'}=\bar k_{\g'}}.
\eeq
The horizontal red dashed line denote  $\dot n_{\theta}=\left. 0.2 \sqrt{Lk} \times  \times \frac{e^2K^2 k_{\gamma'}^4 }{8 \pi  \gamma ^2 k^2 w_{\g'}}\right|_{k_{\g'}=\bar k_{\g'}}$ which agrees with the numerical integration.  

For larger $m_{\g'}$, one cannot have delta function argument close to zero (c.f. Fig.\ref{fig:Multi}) and thus it gets highly suppressed. Therefore there is a resonance like effect at the value \Eq{f'zero} for $\dot{n}_{\theta}$. 

% with the large $\theta\gg 1/\g$ and large $k_{\g'}$ we 
%my estimation is not reliable. We need to have $ \frac{k_{\g'}^{x,y} K}{\g k \b^z}\ll 1 $ for the expansion. 

\label{app:circular}
\bibliography{undulator3.bib}

\end{document}